\documentclass[iop]{emulateapj}

\usepackage{graphicx,longtable}
\usepackage{array}
\usepackage{amsmath}
\usepackage{amssymb}
\usepackage{subfigure}
\usepackage{natbib}
\def\Msun{~M_{\odot} }

\begin{document}
\title{IMPACTS OF ROTATION ON THREE-DIMENSIONAL HYDRODYNAMICS
OF CORE-COLLAPSE SUPERNOVAE}

\author{Ko Nakamura\altaffilmark{1,2}, 
Takami Kuroda\altaffilmark{1,3},
Tomoya Takiwaki\altaffilmark{4}, 
 and Kei Kotake\altaffilmark{1,4,5}}

\affil{$^1$Division of Theoretical Astronomy, National Astronomical Observatory of Japan,
2-21-1 Osawa, Mitaka, Tokyo, 181-8588, Japan}
\affil{$^2$Faculty of Science and Engineering, Waseda University,
Ohkubo 3-4-1, Shinjuku, Tokyo, 169-8555, Japan}
\affil{$^3$Department of Physics, University of Basel,
Klingelbergstrasse 82, 4056 Basel, Switzerland}
\affil{$^4$Center for Computational Astrophysics, National Astronomical Observatory of Japan,
2-21-1 Osawa, Mitaka, Tokyo, 181-8588, Japan}
\affil{$^5$Department of Applied Physics, Fukuoka University,  
8-19-1 Nanakuma Jonan, Fukuoka, 814-0180, Japan}

\begin{abstract}
We perform a series of 
simplified numerical experiments to explore how rotation impacts 
the three-dimensional (3D) hydrodynamics of core-collapse supernovae.
For the sake of our systematic study,
 we employ a light-bulb scheme to trigger explosions
 and a three-flavor neutrino leakage scheme to treat 
deleptonization effects and neutrino losses from proto-neutron star interior.
Using a $15 \Msun$ progenitor, 
we compute thirty models in 3D with a wide variety of 
initial angular momentum and light-bulb neutrino luminosity.
We find that the rotation can help onset of neutrino-driven 
explosions for the models in which the initial angular momentum is 
matched to that obtained in recent stellar evolutionary calculations
($\sim 0.3$-$3$ rad s$^{-1}$ at the center). 
For the models with larger initial angular momentum, 
the shock surface deforms to be more oblate 
due to larger centrifugal force. This makes not only a gain region 
more concentrated around the equatorial plane, but also 
the mass in the gain region bigger.
As a result, buoyant bubbles tend to be coherently formed 
and rise in the equatorial region, which pushes 
the revived shock ever larger radii until a global explosion is triggered. 
We find that these are the main reasons that the preferred 
direction of explosion in 3D rotating models is often perpendicular to 
the spin axis, which is in sharp contrast to the polar explosions 
around the axis that was obtained in 
previous 2D simulations.
\end{abstract}
\medskip

\keywords{hydrodynamics --- neutrinos --- supernovae: general}

\maketitle

\vskip 1.3cm


\section{INTRODUCTION}
Multi-dimensionality in the inner working of core-collapse supernovae 
(CCSNe) has long been considered as one of the most important 
 ingredients to understand the explosion mechanism. 
 Shortly after the discovery of pulsars in the late 1960's 
\citep{hewish}, rotation and magnetic fields were first
proposed to break spherical symmetry of the supernova engine 
\citep{leblanc,bisno76,muller79}.
In the middle of 1980's, the delayed 
neutrino-driven mechanism was proposed by \citet{wils85} and 
\cite{bethe85} who 
 clearly recognized 
the importance of multi-dimensional (multi-D) effects 
\citep{wilson88}\footnote{because they obtained 
more energetic explosions in the 
spherically symmetric (1D) models when 
proto-neutron-star (PNS) convection was phenomenologically taken into 
account.}. 
Ever since SN1987A, 
 multi-D hydrodynamic simulations have been carried out extensively
in a variety of contexts. They give us a confidence that multi-D
 hydrodynamic motions associated with PNS convection
\citep[e.g.,][]{Keil96,mezzacappa98,bruenn04,dessart06}, 
neutrino-driven convection 
\citep[e.g.,][]{herant,burr95,jankamueller96,fryer04a,Murphy08}, and the 
Standing-Accretion-Shock-Instability 
\citep[SASI, e.g.,][and see references therein]{Blondin03,Ohnishi06,Scheck08,thierry,Iwakami08,Iwakami09,Kotake09,rodrigo09,fern13} can help the onset of neutrino-driven 
explosions. In fact, a growing 
number of neutrino-driven models have been recently reported in the
 first-principle two-dimensional (2D) simulations 
(\citealt{Buras06b,Buras06a,Ott08,marek,bruenn,bruenn13,suwa10,suwa13,BMuller12b,BMuller13}, however, \citealt{dolence14}).

This success, however, is bringing to light new questions.
One of the outstanding problems is that the explosion energies 
obtained in these 2D models from first principles (though some of them were reported 
before their explosion energies saturated)
are typically smaller by one 
order of magnitudes to 
explain the canonical supernova kinetic energy 
($\sim 10^{51}$ erg, see Tables in \citet{Kotake13} and \citet{Papish14} for a summary).  
Most researchers are now seeking for some possible physical 
ingredients to make these underpowered explosions more energetic 
\citep[see][for recent reviews]{Janka12,Burrows13,Kotake12_ptep}.

One of the major candidates is three-dimensional (3D) effects on the neutrino-driven 
mechanism. Employing a ``light-bulb'' scheme in 3D simulations,
 \citet{nordhaus10} were the first to point out that 3D leads
 to easier explosions than 2D. This was basically supported by some of the 
follow-up studies \citep{burrows12,dolence12,murphy13}, but not
 by 3D simulations with a similar setup \citep{hanke11,couch12} 
 and by 3D simulations with spectral neutrino transport 
\citep{Takiwaki12,Hanke13,takiwaki14}.
 Another prime candidate is general relativity (GR),
 which has been agreed to help neutrino-driven explosions in both 
2D \citep{BMuller12b,BMuller13} and 3D models 
(\citet{KurodaT12}, see also \citet{Ott12a,kuroda13}). 
The impacts of nuclear equations of state (EOS) have been
investigated in 2D models \citep{marek,Marek09,suwa13,couch13},
and these studies reached an agreement that softer EOS leads to easier explosions.
 The neutrino-driven mechanism could be assisted by 
 the magnetohydrodynamic (MHD) mechanism that works
 most efficiently when the pre-collapse cores have rapid rotation and 
strong magnetic fields 
(e.g., \citet{Kotake06} and \citet{moesta14} for collective references therein). 
It should be mentioned that even in the non-rotating case, 
 the MHD mechanism assists the onset of explosion 
via the magnetic-field amplifications due to the SASI 
\citep{endeve,endeve12,martin11}.
Other possibilities include nuclear 
 burning behind the propagating shock \citep{bruenn06,nakamura12,yamamoto13}, additional heating reactions in the gain region \citep{sumiyoshi08,arcones,furusawa13}, hadron-quark phase transitions \citep[e.g.,][]{sage09,fischer12}, or energy dissipation by the magnetorotational 
instability \citep{thomp05,Obergaulinger09,masada12,sawai13,sawai14}.

Joining in these efforts to look for some possible ingredients to 
foster explosions, we investigated the roles of rotation in this 
study. In 2D simulations of a $15 M_{\odot}$ progenitor 
with detailed neutrino transport, 
 \citet{marek} were the 
first to observe the onset of earlier explosions 
 in models that include moderate rotation
($\Omega_0 = 0.5~{\rm rad}~{\rm s}^{-1}$ with $\Omega_0$ being
 the pre-collapse central angular velocity) than in models 
 without rotation.
By performing 2D simulations of an 11.2 $M_{\sun}$ star 
with more idealized spectral neutrino transport scheme 
\citep[e.g.,][]{idsa},
\citet{suwa10} showed that stronger 
explosion is obtained in models that include rapid rotation 
($\Omega_0 = 2~{\rm rad}$ s$^{-1}$) than those without rotation.
\citet{suwa10} also pointed out that rotation helps explosions, 
not only because the mass enclosed
inside the gain radius becomes larger for rapidly rotating models,
 but also because north-south symmetric bipolar explosions that 
are generally associated with rapidly rotating models can expel much more 
material than that of one-sided unipolar explosions in the non-rotating models. 
These findings naturally open a simple question.
Will these features so far obtained in 2D models also persist 
in 3D models? 

In this study, we performed a series of 
simplified numerical experiments to explore how rotation impacts the 
3D hydrodynamics of the CCSN core that produces an explosion 
 by the neutrino mechanism.
For the sake of our systematic study,
 we employed a light-bulb scheme to trigger explosions \citep[e.g.,][]{jankamueller96,Janka01} and a three-species neutrino leakage scheme 
\citep[e.g.,][]{ross03}
to treat 
deleptonization effects and neutrino losses from the neutron star 
 interior (above an optical depth of about unity).
It is well-known from interferometric observations 
\citep[see][for a review]{vanbelle} 
that stars more massive than 1.5 $M_{\odot}$ are generally rapid rotators 
\citep{huang06,huang08}. 
However, due to numerical difficulties of multi-D stellar evolutionary
 calculations (see \citet{maeder12} for a review, and
\citet{meakin07,meakin2011} for recent developments), it is quite 
uncertain  
how such high surface velocities are entirely evolved 
during stellar evolution till the onset of core-collapse, in which
 multi-D hydrodynamics including mass-loss, rotational 
mixing, and magnetic braking is playing an active role in determining the 
 angular momentum transport.
In this study, we made pre-collapse models by parametrically
 adding the initial angular momentum to a widely used a
15 $M_{\odot}$ progenitor \citep{WW95}.
We carried out 3D special-relativistic simulations
starting from the onset of gravitational 
collapse, through bounce, trending towards explosions (typically 
up to about $\sim $1 s postbounce) and compare results of 
thirty 3D models, in which 
the input neutrino luminosity and the initial rotation rate 
are systematically varied.
 We found that a critical neutrino luminosity to 
obtain neutrino-driven explosions becomes generally smaller 
for models with larger initial angular momentum. Our 
3D models show a much wider variety of the explosion geometry than in 2D. 
In the 3D rotating models, we found that 
the preferred direction of explosion is often {\it perpendicular} to the spin axis, which 
 is in sharp contrast to polar explosions around the 
 spin axis that was commonly obtained in previous 2D simulations.

We begin in Section \ref{sec:Numerical Method} with a description of numerical setup and 
 initial models. The main results are shown in Section \ref{sec:results}.
We summarize our results and discuss their implications in Section \ref{sec:conclusion}.

\section{NUMERICAL SETUP}
\label{sec:Numerical Method}

By utilizing the code developed by \cite{KurodaT10} and \citet{KurodaT12}, 
we perform 3D, special-relativistic (SR) hydrodynamic simulations of 
the collapse, bounce, and post-bounce evolution of the core of 
massive stars. 
The basic evolution equations are written in a conservative form as,
\begin{eqnarray}
\label{eq:SRmass}
\partial_t \rho_{\ast}+\partial_i(\rho_\ast v^i)&=&0,\\
\label{eq:SRmomentum}
\partial_t  S_i+\partial_j( S_i v^j+P\delta_i^j)&=&-\rho_\ast\partial_i\phi-Qu_{i}, \\
\label{eq:SRenergy}
\partial_t  \tau+\partial_i ( \tau v^i+P v^i)&=&-\rho_\ast v^i \partial_i\phi-QW,\\
\label{eq:SRlepton}
\partial_t (\rho_\ast Y_e)+\partial_i (\rho_\ast Y_e v^i)&=&\rho_\ast \Gamma_e,
\end{eqnarray}
 where $\rho_\ast\equiv\rho W$, 
$S_i\equiv\rho hW u_i $, and $S_0 \equiv\rho h W^2-P$ are 
auxiliary variables (corresponding to density, momentum, and energy),
 $\rho$ is the rest mass density, $W$ the Lorentz factor, $u_\mu$ the 4-velocity of fluid,
$h\equiv 1+\varepsilon+P/\rho$ the specific enthalpy,
$v^i=u^i/u^0$, $\tau = S_0 - \rho_{\ast}$, $Y_e$ the electron fraction,
$\varepsilon$ and $P$ the internal energy and pressure, 
$\delta_i^j$ the Kronecker delta, respectively.
We employ the EOS based on the relativistic 
mean-field theory of \citet{Shen98}.
As pointed out by \citet{suwa13} and \citet{couch13}, 
the use of Lattimer-Swesty EOS 
with an incompressibility modulus of $K = 220$ MeV (LS220), 
which is another representative EOS employed in modern CCSN simulations, 
would lead to easier explosion because LS220 EOS is softer than Shen EOS. 
Detailed analysis to clarify the impacts of EOS in rotating core-collapse 
models remains to be studied in the future work.

In the right-hand-side of above equations, $\phi$ represents 
gravitational potential, $Q$ and $\Gamma_e$ denote exchange of 
energy-momentum and lepton number between neutrino and matter.
To solve the Poisson equation of self-gravity ($\nabla^2\phi=4\pi S_0$),
we employ a 
``BiConjugate Gradient Stabilized (BiCGSTAB)'' method 
with an appropriate boundary condition 
(see \citet{KurodaT12} for further details). 

The neutrino-matter interaction term $Q$ is
 divided into cooling and heating terms ($Q = Q^C + Q^H$).
Following a methodology of a multi-flavor 
neutrino leakage scheme 
\citep[e.g.,][]{ross03},
 the cooling term can be estimated as,
\begin{eqnarray}
&Q^C&=Q_{e^-}^{f}+Q_{e^-}^{h}+Q_{e^+}^{f}+Q_{e^+}^{h}\nonumber \\
&+&\sum_{\nu\in(\nu_e,\bar{\nu}_e,\nu_x)}
2(Q_{e^-e^+\rightarrow \nu\bar\nu}+Q_{\gamma\rightarrow \nu\bar\nu}+Q_{NN\rightarrow NN\nu\bar\nu}),
\end{eqnarray}
 where $Q_{e^{-/+}}^{f}$ and $Q_{e^{-/+}}^{h}$ represents the cooling rate
 by electron/positron capture on free nucleons and on heavy nuclei, 
 $Q_{e^-e^+\rightarrow \nu\bar\nu}$, $Q_{\gamma\rightarrow \nu\bar\nu}$,
 and $Q_{NN \rightarrow NN \nu\bar\nu}$ denote contributions
from pair neutrino annihilation, plasmon decay, and 
 nucleon-nucleon bremsstrahlung, 
respectively 
(see \citet{KurodaT12} for more detail).
Following \citet{Janka01} and \citet{Murphy08}, we employ a simple light-bulb scheme
 to estimate $Q^H$ as,
\begin{eqnarray}
Q^H =  1.544 \times 10^{20} \left( \frac{L_{\nu_{\rm e}}}{10^{52} \, {\rm erg \, s^{-1}}}\right)
 \left( \frac{T_{\nu_e}}{4{\rm MeV}} \right)^2 \nonumber \\
\times \left( \frac{r}{100 {\rm km}} \right)^{-2}
(Y_{\rm n}+Y_{\rm p}) e^{-\tau} {\rm [erg/g\,s]},
\end{eqnarray}
\label{eq:nheat}
where $L_{\nu_{\rm e}}$ is the electron-neutrino luminosity 
 that is 
assumed to be equal to the anti-electron neutrino luminosity 
($L_{\bar{\nu}_{\rm e}} = L_{\nu_{\rm e}}$), $T_{\nu_{\rm e}}$ is 
the electron neutrino temperature assumed to be kept constant as $4$ MeV, $r$ 
is the distance from the center, 
$Y_n$ and $Y_p$
are the neutron and proton fractions, and
$\tau$ 
is the electron neutrino
optical depth that we estimate from Equation (7) in \citet{hanke11}.

As already mentioned, we add pre-collapse rotation
to the non-rotating 15$M_\odot$ model
\citep[model ``s15s7b2'']{WW95} to study its effect in a controlled fashion.
We assume a shell-type rotation profile as 
\begin{equation}
\Omega(r) = \Omega_0 \frac{R_0^2}{r^2 + R_0^2},
\label{eq:rot}
\end{equation}
where $\Omega(r)$ is the angular velocity at the radius of $r$, 
$R_0$ set here to be $2\times 10^8$ cm is reconciled
with results from stellar evolution calculations suggesting uniform 
rotation in the pre-collapse core. 
The initial angular velocity at the origin, $\Omega_0$, is 
treated as a free parameter and we vary it as 
$\Omega_0=0,\ 0.1\,\pi,$ and $0.5\,\pi$ rad s$^{-1}$. Note that these angular velocities are in good agreement 
 with the outcomes of most recent stellar evolution models that
 were evolved with the inclusion of magnetic fields 
\citep[$\Omega_0 \sim 0.2$-$0.3$ rad s$^{-1}$ of models m15b4 and m20b4 in][]{Heger05} 
or without the magnetic braking 
\citep[$\Omega_ 0 \sim 3$-$4 $ rad s$^{-1}$ of models E15 and E20 in][]{Heger00a}.

The 3D computational domain consists of a cube of $10000^3~{\rm km}^3$
volume in the Cartesian coordinates. Nested boxes 
with nine refinement levels are embedded in the computation domain. 
We use a fiducial resolution at the maximum 
refinement level (the minimum grid scale) 
of $\Delta x_{\rm min} = 600$ m in each direction. 
Each block has $64^3$ cubic cells. Roughly speaking, such structure 
corresponds to an 
angular resolution of $\sim 2$ deg through the entire computational 
domain. To study the resolution dependence of our results, we perform high 
resolution runs for some selected models, in which each nested block has 
$128^3$ cubic cells 
(i.e. $\Delta x_{\rm min} = 300$ m at the centre). 
Note that 
the numerical resolutions in the central region 
are almost similar or slightly better comparing to 
recent 3D Newtonian simulations with a similar treatment of the neutrino effects 
\citep[$\Delta x_{\rm min} = 700$ m of][]{couch12}, 
as well as GR simulations 
\citep[$\Delta x_{\rm min} = 370$ m of][]{Ott12a}, 
although \citet{Ott12a} adopted the leakage scheme for both heating and cooling.

\begin{table*}[htbp*]
\begin{center}
\caption{Model Summary}
\begin{tabular}{cccccccccccc}
\hline \hline
                            & \multicolumn{3}{c}{$\Omega_0 = 0.0$ (rad s$^{-1}$)} &
                            & \multicolumn{3}{c}{$\Omega_0 = 0.1\,\pi$ (rad s$^{-1}$)}&
                            & \multicolumn{3}{c}{$\Omega_0 = 0.5\,\pi$ (rad s$^{-1}$)}\\
                            \cline{2-4} \cline{6-8} \cline{10-12}
${L_{\nu,52}}^a$     & ${t_{\rm exp}}^b$  & $\dot{M}$$^c$ & shape$^d$&
                                & $t_{\rm exp}$        & $\dot{M}$ & shape      &
                                & $t_{\rm exp}$        & $\dot{M}$ & shape      \\
($10^{52} \, {\rm erg \, s}^{-1}$)        
                                   & (ms)&($\Msun \, {\rm s}^{-1}$)&    &
                                   & (ms)&($\Msun \, {\rm s}^{-1}$)&    &
                                   &(ms)&($\Msun \, {\rm s}^{-1}$)&    \\
\hline
1.5                       & & & && ---    & ---     & ---       && 135 & 0.426 & oblate \\
1.7                       & & & && 541 & 0.263 & oblate   && 128 & 0.469 & oblate \\
1.9                       & & & && 500 & 0.264 & oblate   && 117 & 0.579 & oblate \\
2.1                       & & & && 355 & 0.293 & oblate   && 108 & 0.679 & oblate \\
2.3                       & ---  & ---  &---        && 336 & 0.300 & oblate  && 90 & 0.907 & oblate \\
                            &       &       &              &&563$^{*,e}$ & 0.264$^{*}$ & oblate$^{*}$ &&  &  &   \\
2.5                       & ---  & ---  &---         && 348 & 0.295 & oblate  && 93 & 0.864 & oblate \\
                            &       &       &              && 203$^{*}$ & 0.356$^{*}$ & spher.$^{*}$ &&  &   &    \\
2.7                       & ---  & ---   &---          && 229 & 0.345 & oblate  && 70 & 1.09 & oblate \\
                            &       &        &               &&128$^{*}$ & 0.469$^{*}$ & spher.$^{*}$ &&    &    &  \\
2.9                       & 207 & 0.355 & spher.      &&   57 &   1.20 & spher.  &&        &         &\\
                            &     &    &   && 75$^{*}$ & 1.05$^{*}$ & spher.$^{*}$ &&               & &\\
3.1                       & 168 & 0.369 & spher.      &&&&   &&&&\\
3.3                       & 140 & 0.404 & spher.        &&&&   &&&&\\
3.5                       & 47 & 1.31 & spher.            &&&&   &&&&\\
\hline \hline
\multicolumn{12}{l}{$^{a}$ Electron-neutrino luminosity. }\\
\multicolumn{12}{l}{$^{b}$ Postbounce time of onset of explosion. 
Models that have a blank entry in the table are}\\
\multicolumn{12}{l}{not simulated and the ``---'' symbol indicates that the model does not explode.}\\
\multicolumn{12}{l}{   during the simulated period of evolution.}\\
\multicolumn{12}{l}{$^{c}$ Mass accretion rate at onset of explosion.}\\
\multicolumn{12}{l}{$^{d}$ Morphological classification of shock shape at $t=t_{\rm exp}$. 
See the text for definition.}\\
\multicolumn{12}{l}{$^{e}$ The values with an asterisk symbol ``xxx$^{*}$'' 
indicates that 
the values are obtained in the higher}\\
\multicolumn{12}{l}{resolution model ($\Delta x_{\rm min} = 300$m) 
compared to our fiducial resolution of $\Delta x_{\rm min} = 600$m.}\\

\\
\\
\end{tabular}
\label{tbl-summary}
\end{center}
\end{table*}

\section{RESULTS}
\label{sec:results}

\subsection{Overview of Simulation Results}\label{overview}

Following the previous investigations 
\citep[e.g.,][]{Murphy08,nordhaus10,hanke11,couch12}, 
we first take the approach of \citet{goshy}, 
using critical neutrino luminosity 
versus accretion rate 
to overview the impacts of rotation.
Hereafter the neutrino luminosity is referred to as 
$L_{\nu, 52}$ in unit of $10^{52}$ erg s$^{-1}$, 
and the initial angular velocity $\Omega_0$ in Equation (\ref{eq:rot}) is 
given in unit of rad s$^{-1}$ 
unless otherwise noted.

For three cases of rotation ($\Omega_0 = 0.0$, $0.1\,\pi$, and $0.5\,\pi$), 
Table \ref{tbl-summary} summarizes the time of explosion, $t_{\rm exp}$, 
the mass accretion rate, $\dot{M}$, 
and the shape of shock front,  
as a function of 
$L_{\nu, 52}$. 
Here as in \citet{nordhaus10} and \citet{hanke11}, we define $t_{\rm exp}$ 
as the time when the shock reaches an average radius 
of 400 km (and does not returns later on). 
$\dot{M}$ is estimated at $t=t_{\rm exp}$ and $r=500$ km. 
The shock configuration is defined by the ratio of shock radius 
around the equatorial plane to that in the polar directions at $t_{\rm exp}$. 
The shock morphology is discussed in more detail in the last of \S \ref{overview}.

Given an input luminosity, 
for example $L_{\nu, 52}=2.5$, 
the non-rotating model ($\Omega_0 = 0.0$) does not explode 
till 1 second after bounce (so the model is labeled as ``---''). 
On the other hand, the corresponding model with moderate rotation 
($\Omega_0 = 0.1\,\pi$) explodes 
at $t_{\rm exp}=348$ ms postbounce and the rapidly 
rotating model 
($\Omega_0 = 0.5\,\pi$)
 explodes much faster (at $t_{\rm exp}=93$ ms).
This is a common trend for models with the input luminosity 
 below $L_{\nu, 52}= 2.9$. 
The critical value of the neutrino luminosity and accretion rate of the non-rotating model 
is $L_{\nu, 52}= 2.9$ and $\dot{M} \sim 0.36 \Msun~{\rm s}^{-1}$ 
(Table \ref{tbl-summary}).
Note that these values for the critical neutrino luminosity and the critical
 mass accretion rate in the non-rotating model
are roughly in agreement with those 
obtained 
in previous 3D simulations with the light-bulb scheme
(e.g., $L_{\nu, 52}=2.5$ in \citet{hanke11} and $L_{\nu, 52}=2.1$ in 
\citet{couch12} for $\dot{M} \sim 0.3 \Msun~{\rm s}^{-1}$)\footnote{
The difference from the previous models should come from deleptonization and cooling inside the PNS,
 which was not treated in the previous light-bulb models, but 
is treated by the leakage scheme in this study. The inclusion of 
deleptonization leads to smaller $Y_e$ in the PNS, which makes the PNS 
more compact and the gravitational pull from the PNS stronger, 
which is most likely to explain the reason of the higher critical luminosity 
in this study.}.

\begin{figure*}[htbp]
\begin{center}
\includegraphics{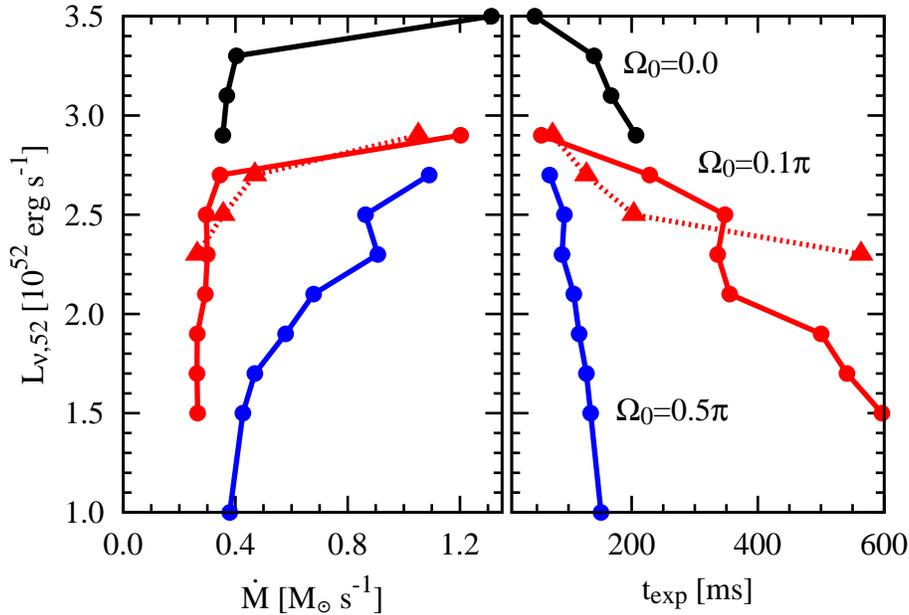}
\caption{Critical curves for the 
neutrino luminosity $L_{\nu,52}$
versus mass accretion rate ($\dot{M}$, left plot) and
 versus explosion time ($t_{\rm exp}$, right plot) for simulations with $\Omega_0 = 0.0$ (black), $0.1\,\pi$ (red), and $0.5\,\pi$ rad (blue).
The time of explosion is defined at the moment when the average shock radius reaches 400 km, and accretion rate is measured at the distance of 500 km from the center at the time of $t_{\rm exp}$.
Two cases of spatial resolution are shown in each panel, 
the fiducial resolution ($\Delta x_{\rm min} = 600$m) by solid lines with
 filled circles and
 the high resolution ($\Delta x_{\rm min} = 300$m) by dashed lines with filled triangles.}
 \label{fig-summary}
\end{center}
\end{figure*}

The left panel of Figure \ref{fig-summary} shows the $L_{\nu,52}$-$\dot{M}$ 
curves for all the computed models with different initial rotation
 rates of $\Omega_0 = 0.5\,\pi$ (blue line),
 $0.1\,\pi$ (red line), and
 $0.0$ (black line).
Compared to the non-rotating models (black line),
the critical luminosities are shown to be smaller 
by $\sim 50 \%$ for the mildly rotating models (red line) and 
by $\sim 70 \%$ for the rapidly rotating models (blue line). 
These results demonstrate that rotation 
could result in easier explosions 
compared to the non-rotating models. 
Very recently, 
\citet{Iwakami14} obtained qualitatively the same results with an ideal initial setup.

Not surprisingly, a high driving luminosity 
is necessary to overcome a high mass accretion rate regardless 
of rotation (left panel), which results in earlier onset of explosion 
(right panel). 
These features are qualitatively in accord with those obtained 
in previous parameterized 3D models without rotation 
\citep[e.g.,][]{burrows12,dolence12,hanke11}. 
Here it should be noted that the initial angular momentum 
of our moderately-rotating models ($\Omega_0 = 0.1\,\pi$)
is not as rapid as previously assumed \citep{marek,suwa10},
 but as slow as that predicted by the most recent stellar evolution 
calculation \citep{Heger05}. As we discuss more in detail 
 in the following, even such a slow rotation can significantly effect the 
shock revival and evolution in the postbounce phase.

\begin{figure}[tbp]
\begin{center}
\includegraphics[scale=0.7]{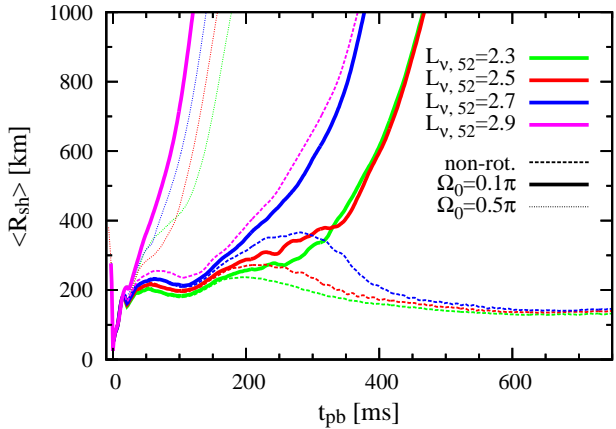}
\caption{Time evolution of the average shock radius as a function of 
the postbounce time $t_{\rm pb}$ for 3D models with rapid rotation 
($\Omega_0 = 0.5\,\pi$, dotted lines), moderate rotation 
($\Omega_0 = 0.1\,\pi$, solid lines), and 
no rotation (dashed lines). Different electron-neutrino luminosity 
(in unit of 
$10^{52}~{\rm erg}~{\rm s}^{-1}$, $L_{\nu, 52}$) 
is plotted in different colors.}
\label{fig-rsh}
\end{center}
\end{figure}

Figure \ref{fig-rsh} shows time evolution of average shock radius 
$\langle R_{\rm sh} \rangle$ for some selected luminosity 
models ($L_{\nu, 52}$ between 2.3 and 2.9 presented in different colors)
 with the different rotation parameters
($\Omega_0 = 0$ in dashed lines, 
$\Omega_0 = 0.1\,\pi$ in solid lines, 
and $\Omega_0 = 0.5\,\pi$ in thin dotted lines). 
Without rotation (dashed lines), one can see that  
the bounce shock all stalls at the radius of $r \sim 200$ km as 
 in previous 3D models with similar microphysical setup \citep {hanke11,dolence12}. 
As is expected, the maximum shock extent afterwards is bigger for models with
 higher input luminosity 
(i.e., biggest for the dashed pink line ($L_{\nu, 52} = 2.9$),
 followed in order by the dashed blue ($L_{\nu, 52} = 2.7$), 
dashed red ($L_{\nu, 52} = 2.5$), and dashed green line ($L_{\nu, 52} = 2.3$)).
Except for the highest luminosity model (with $L_{\nu, 52}  = 2.9$, dashed pink
 line), the late-time shock trajectory of the non-rotating models
 similarly shows a continuous recession and the stalled shock never revives
 during the simulation time.

When these models have moderate rotation initially 
($\Omega_0 = 0.1\,\pi$, solid lines), the shock trajectories
 show a clear deviation from those without rotation. This occurs
 typically after $\sim 200$ ms postbounce (see the bifurcation points
 between the solid and dashed lines for models with $L_{\nu, 52} =$ 2.3 (green), 2.5 (red), 
and 2.7 (blue). This clearly demonstrates that even the moderate pre-collapse 
rotation could effect the shock evolution. 
When the initial rotational speed is very fast
 ($\Omega_0 = 0.5\,\pi$, thin dotted lines), the shock revival
 occurs very quickly after bounce even for the model with the 
lowest luminosity.
We will address the reasons of such interesting behaviors from the next secsion. 
Before going into detail,  we briefly summarize how rotation 
impacts the 3D blast morphology in the following.

\begin{figure*}[htbp]
\begin{center}
\includegraphics[scale=0.8]{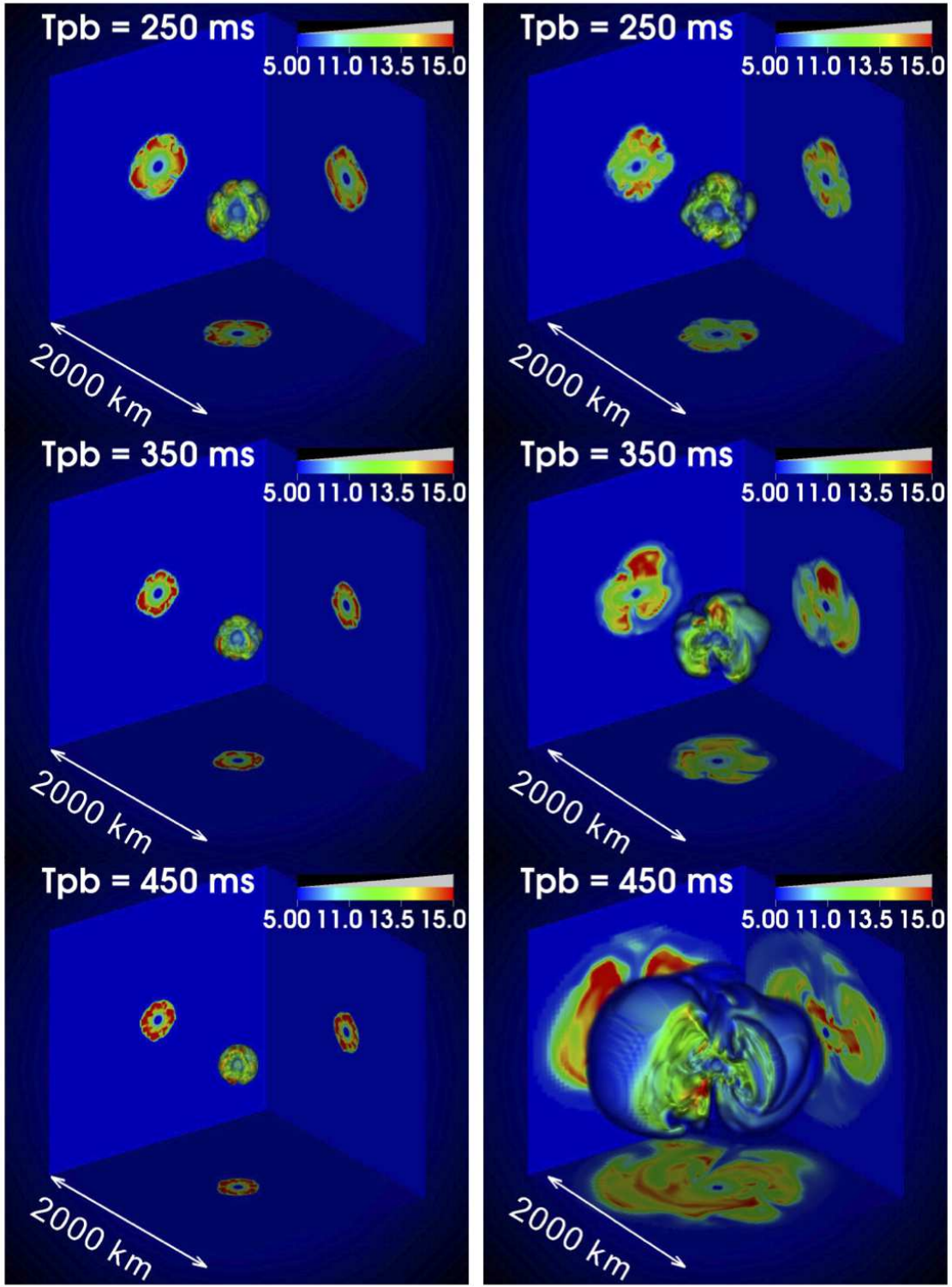}
\caption{Snapshots of 3D entropy distribution in the fiducial resolution. 
The non-rotating ($\Omega_0=0.0$, left panels) 
and moderately rotating ($\Omega_0=0.1\,\pi$, right panels) models 
are computed with the same input luminosity of $L_{\nu, 52} = 2.5$. Top, middle, 
and bottom panel corresponds to the
 timeslice of 
 $t_{\rm pb}=250$ ms, 350 ms, and 450 ms postbounce,
 respectively.  The vertical axis is aligned with the rotational axis.
Note that spatial scale is different between the left ($600^3$ km$^3$) and right ($2000^3$ km$^3$) panels.}
\label{fig-snap25}
\end{center}
\end{figure*}

 Figure \ref{fig-snap25} shows several snapshots of 3D entropy distribution   
of the $L_{\nu, 52}=2.5$ model 
without rotation ($\Omega_0 = 0.0$, left panels) and 
with moderate rotation ($0.1\,\pi$, right) 
at $t_{\rm pb}=250$ ms (top), 350 ms (middle), and 450 ms 
(bottom), respectively.
After the bounce shock stalls (top left panel),
the shape of the shock keeps to be roundish all the way
for the non-rotating model (from top left to bottom left),
 which does not
 trend towards explosion during the simulation time 
 (see also red dashed line in Figure \ref{fig-rsh}). 
 If the input luminosity is taken to be higher for the 
 non-rotating model, we observe a strong correlation, as previously identified,
 between large-scale  structures; high entropy plumes 
(red regions in the sidewall panels) 
are associated with outflows whereas low-entropy regions 
(blueish regions) are associated with inflows, a natural outcome
 of buoyancy-driven convection \citep{murphy13,burrows12,dolence12,hanke11}.
These postshock structures are clearly different from the same 
 luminosity model but when moderate rotation is initially imposed
($\Omega_0 = 0.1\,\pi$, right panels in Figure \ref{fig-snap25}). An important 
difference between the non-rotating and rotating model 
is the existence of more big-scale structure in the flow of
 the rotating model, which can be seen as an oblate structure in 
the postshock regions (e.g., green regions in the middle right panel; taken
 at $t_{\rm pb} = 350$ ms 
 when the shock reaches to the average radius of 400 km). 
 The natural outcome is an oblate explosion as shown in the bottom right panel.

\begin{figure*}[htbp]
\begin{center}
\includegraphics[scale=0.8]{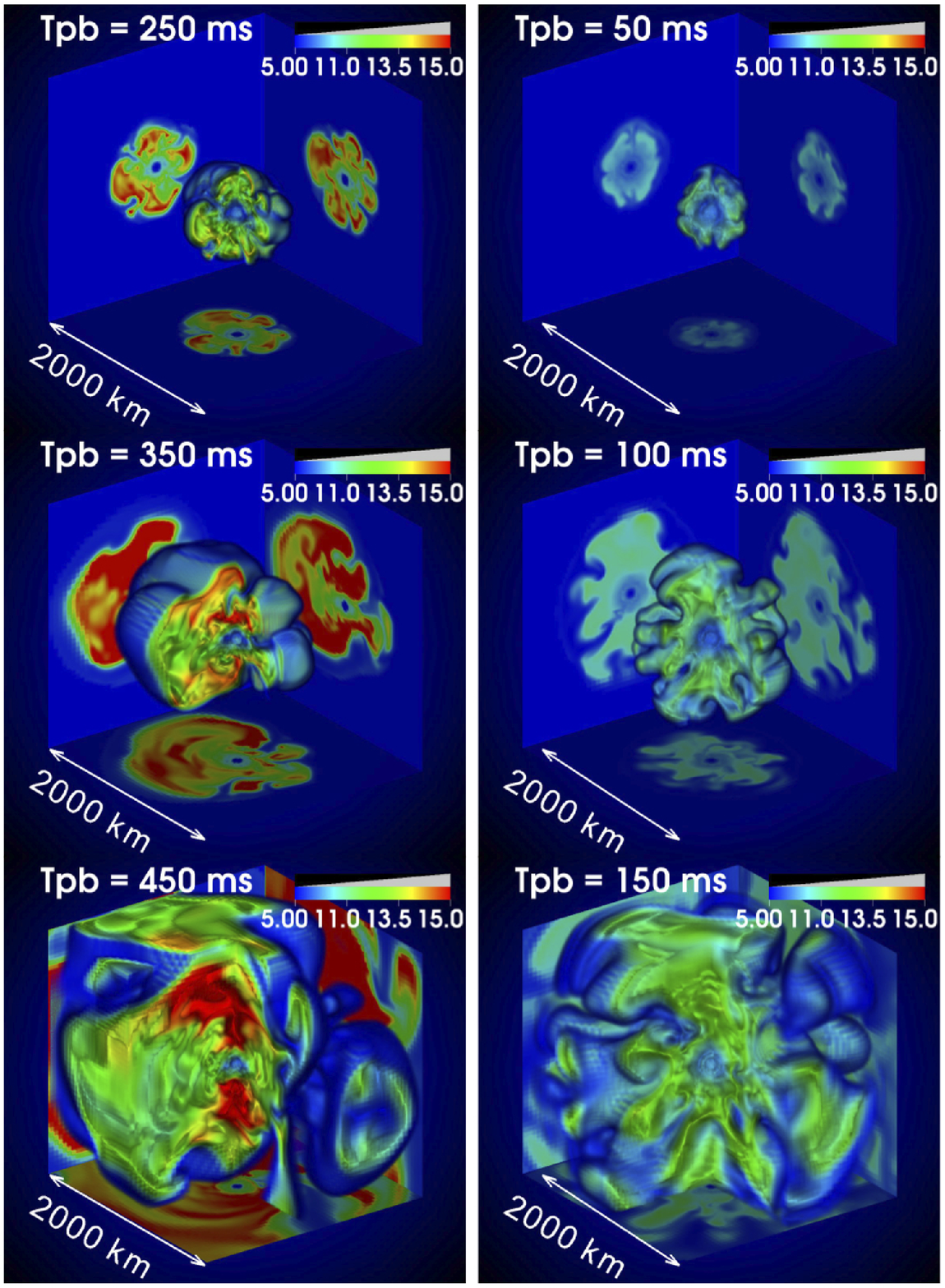}
\caption{Same as Figure \ref{fig-snap25} but for different input 
 luminosity of $L_{\nu, 52} = 2.7$ (left panels) and $2.9$ (right panels) 
with the same initial rotation rate 
(moderate rotation; $\Omega_0=0.1\,\pi$).}
\label{fig-snap2729}
\end{center}
\end{figure*}

The morphology of the revived shock is diverse from models to models.
A snapshot presented in the top left panel of Figure \ref{fig-snap2729} 
(model $L_{\nu,52} = 2.7$ with $\Omega_0 = 0.1\,\pi$)
shows a clear rotational flattening in the postshock region, 
which is a typical feature caused by $l=2$ mode in the rotating core collapse 
and bounce. 
The revived shock of this model (red regions in the middle left panel) 
is also deformed to be oblate. The oblately deformed shock 
moves out triaxially afterwards, leading to a strong shock expansion 
to the equatorial direction rather in an one-sided way 
(bottom left panel). 
For the same initial rotation rate, 
the oblateness is shown 
 to appear more weakly in the model with higher
 input luminosity ($L_{\nu,52} = 2.9$, right panels in Figure \ref{fig-snap2729}).
 This is because in models with higher luminosity the violent fluid motions 
 as well as vigorous convective overturns work to smear out the effects 
 of rotation 
\citep[e.g.,][]{kotake11}. 
This trend is clearly seen in Table \ref{tbl-summary}, 
where the degree of shock deformation is evaluated as 
the ratio of shock radius averaged over $60^\circ \leq \theta \leq 120^\circ$
to that around the rotation axis ($\theta < 60^\circ$, $120^\circ < \theta$).
We define shocks with this ratio between $0.45$ and $0.55$ as ${\it spherical}$, 
and shocks with this ratio superior/inferior to this range is referred to as ${\it oblate/prolate}$.

\begin{figure*}[htbp]
\begin{center}
\includegraphics[scale=0.8]{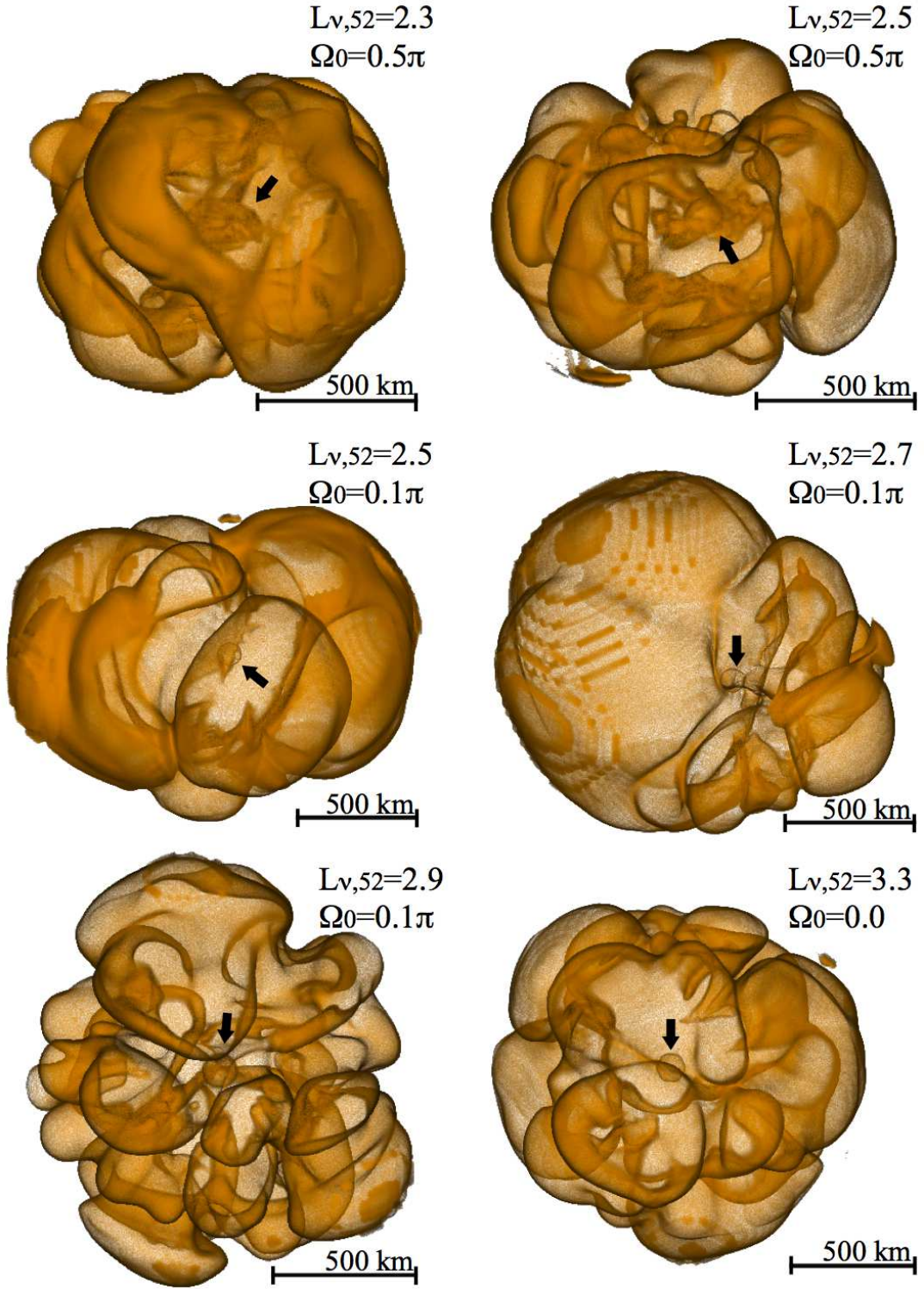}
\caption{Various structures of shocked gas in our models. 
Shown are the isentropic surface of $8 \, k_{\rm B}$ per baryon 
shortly after the shock revival. 
Neutrino luminosity and angular velocity parameters are labeled in each panel.
A obscure sphere pointed by an arrow in each panel corresponds to the 
position of PNSs 
\citep[defined by a fiducial density of $10^{11}~{\rm g} {\rm cm}^{-3}$ as in][]{BMuller12b,suwa13}.
}
\label{fig-isent}
\end{center}
\end{figure*}

To focus on the shock deformation, constant entropy contours 
are shown in Figure \ref{fig-isent}. By comparing with 
 Figures \ref{fig-snap25} and \ref{fig-snap2729} with Figure \ref{fig-isent}, 
 one could more clearly see that 
the blast morphology is close to be 
oblate (top panels and middle left panel in Figure \ref{fig-isent}),
 one-sided (middle
 right panel), and multi-polar (roundish, bottom panels), respectively.
As mentioned above, the high neutrino luminosity (e.g.,
 bottom panels in Figure \ref{fig-isent}) tends to wash away the impacts 
of rotation, so that we mainly focus on models with smaller luminosity
 ($L_{\nu, 52} \leq 2.9$) in the following.

Before going to the next section, we shall note that the above
 results are qualitatively little affected 
by the employed numerical resolution.
 Then what about the quantitative dependence?
 Due to the computational expense, we can currently afford to run four 
high-resolution models in 3D (that we chose to have the canonical initial rotation rate, 
$\Omega_0 = 0.1$). 
As in \citet{hanke11}, our high-resolution models lead to 
 more delayed explosions compared to the low-resolution counterpart
 when the input luminosity is as low as that in \citet{hanke11} 
(see our lowest luminosity model with $L_{\nu, 52} = 2.3$ 
in the right panel of Figure \ref{fig-summary} (dashed red curve) which shows 
the (diagnostic-)explosion time of $t_{\rm exp} \sim 600$ ms). 
On the other hand, our models, switching to high-resolution, 
show earlier explosions ($t_{\rm exp} \lesssim 200$ ms postbounce) 
 when the input luminosity is relatively higher
 ($L_{\nu, 52} \gtrsim 2.5$)\footnote{Recent 3D models
 by \citet{handy} show the similar trend.}. 
We would speculate that high numerical resolution could capture more 
efficiently the growth of such highly convection-dominated explosions
in the case of such high luminosity.
Further numerical experiments are apparently needed
 (at least to get a convergence in 3D), which unfortunately 
demands us a formidable computational cost at present.

\subsection{Detailed Comparison between Non-rotating and Rotating Models}
In this section, we move on to look in more detail into the reason why 
 rotation could result in easier explosion. 
As will be shown, the key is the increase of the total gain mass 
for models with rotation, which is a natural outcome of the rotational 
flattening of the postshock region. 
In the following, we use 
the $L_{\nu,52} = 2.5$ models as a reference, for which the shock revival is 
not obtained in the non-rotating model but is observed 
when moderate rotation ($\Omega_0 = 0.1~\pi$) 
is taken into account (see red lines in Figure \ref{fig-rsh}). 

\subsubsection{Gain Mass}

\begin{figure*}[t]
\epsscale{0.9}
\plottwo{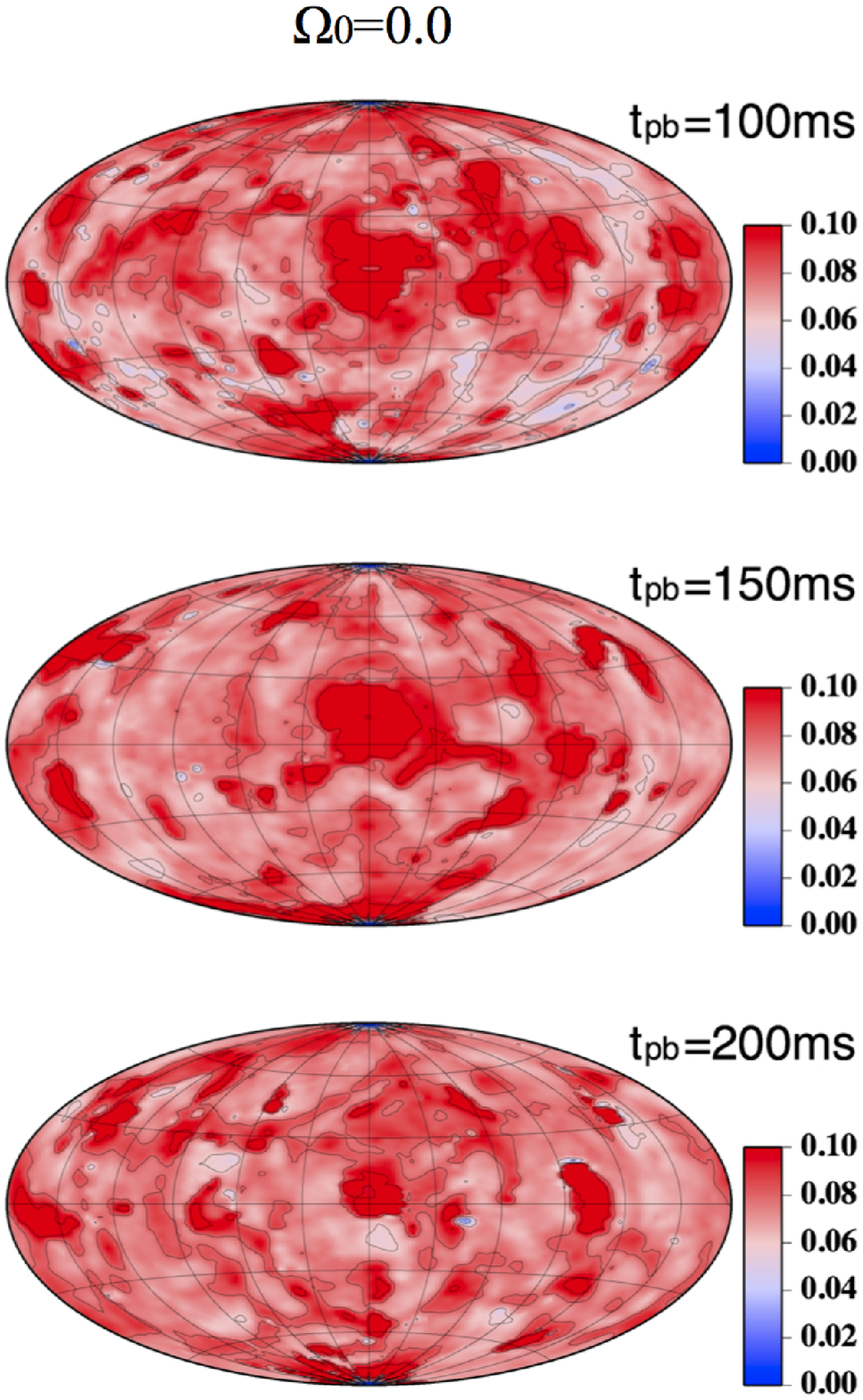}{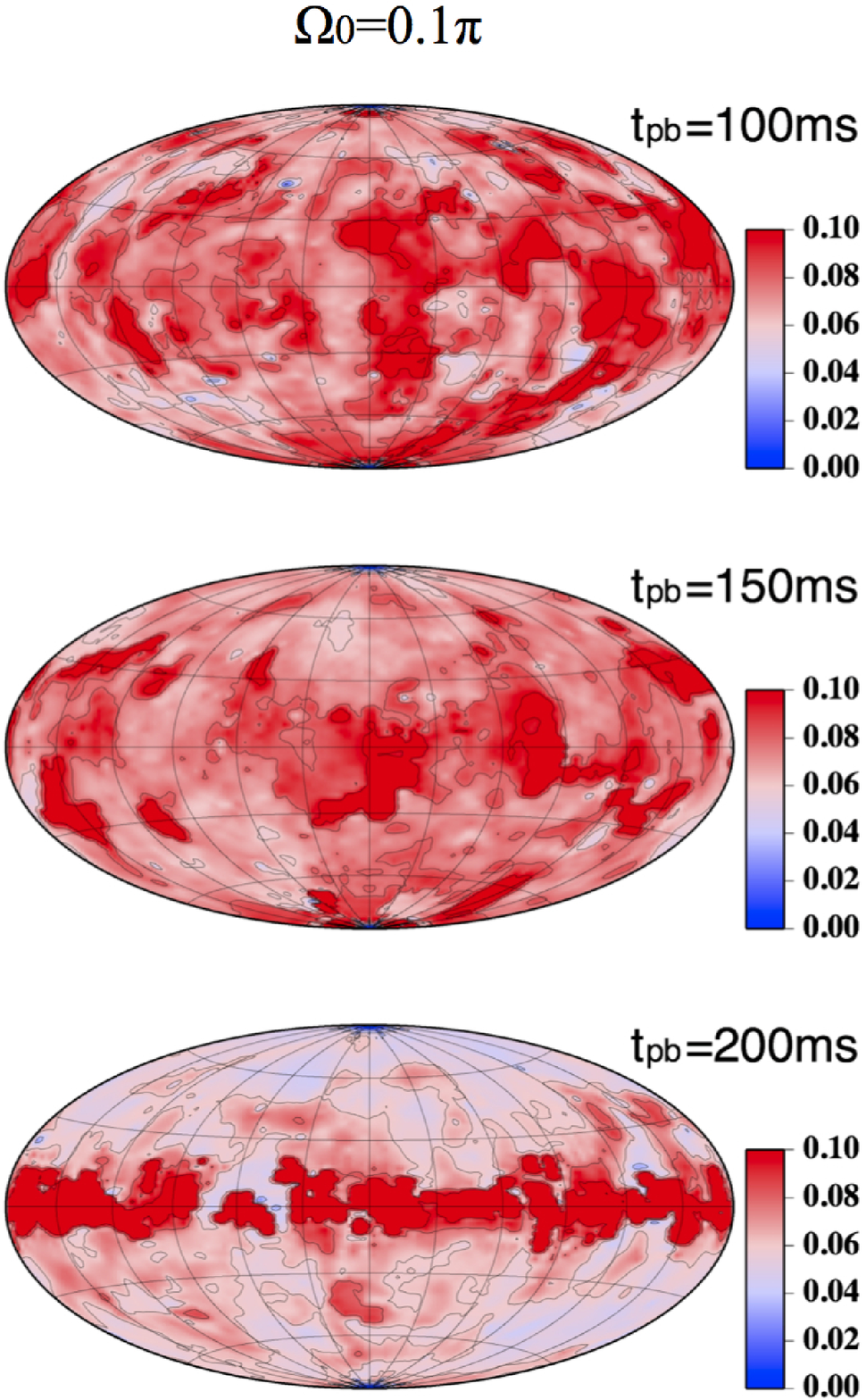}
\caption{Mollweide maps showing the mass distributions in the 
gain region for the $L_{\nu,52} = 2.5$ model with $\Omega_0 = 0.0$ (left) 
and $0.1\,\pi$ (right) at $t_{\rm pb}=100$ ms, 150 ms, and 200 ms 
postbounce (from top to bottom), respectively. 
The color scale shows the mass 
fraction of the gain region in each solid angle (normalized by 
the total mass in the gain region).}  
\label{fig-mlwmg25}
\end{figure*}

It is well known that the enclosed mass in the gain region (the so-called 
gain mass), if bigger (smaller), 
does good (harm) to the revival of the stalled bounce shock.
As a guide to see how rotation affects the gain mass and its spatial 
distributions, 
we plot in Figure \ref{fig-mlwmg25} the Mollweide maps of the gain-mass distributions 
for the $L_{\nu,52} = 2.5$ model without (left panels) and 
with rotation (right panels), respectively.
Throughout our simulation time, the spatial distribution of the 
gain mass stochastically changes (as anticipated) for the non-rotating model
(left panels, see red regions), whereas a clear excess 
is seen for the rotating model in the vicinity 
of the equatorial plane (seen like a belt, bottom right panel). 
The excess is clearly seen typically after $\sim$ 200 ms postbounce,
but a gradual concentration to the equatorial belt (red regions in 
 the equatorial region) can be also seen 
at 150 ms postbounce (middle right panel).

\begin{figure}[tbp]
\begin{center}
\includegraphics[scale=0.5]{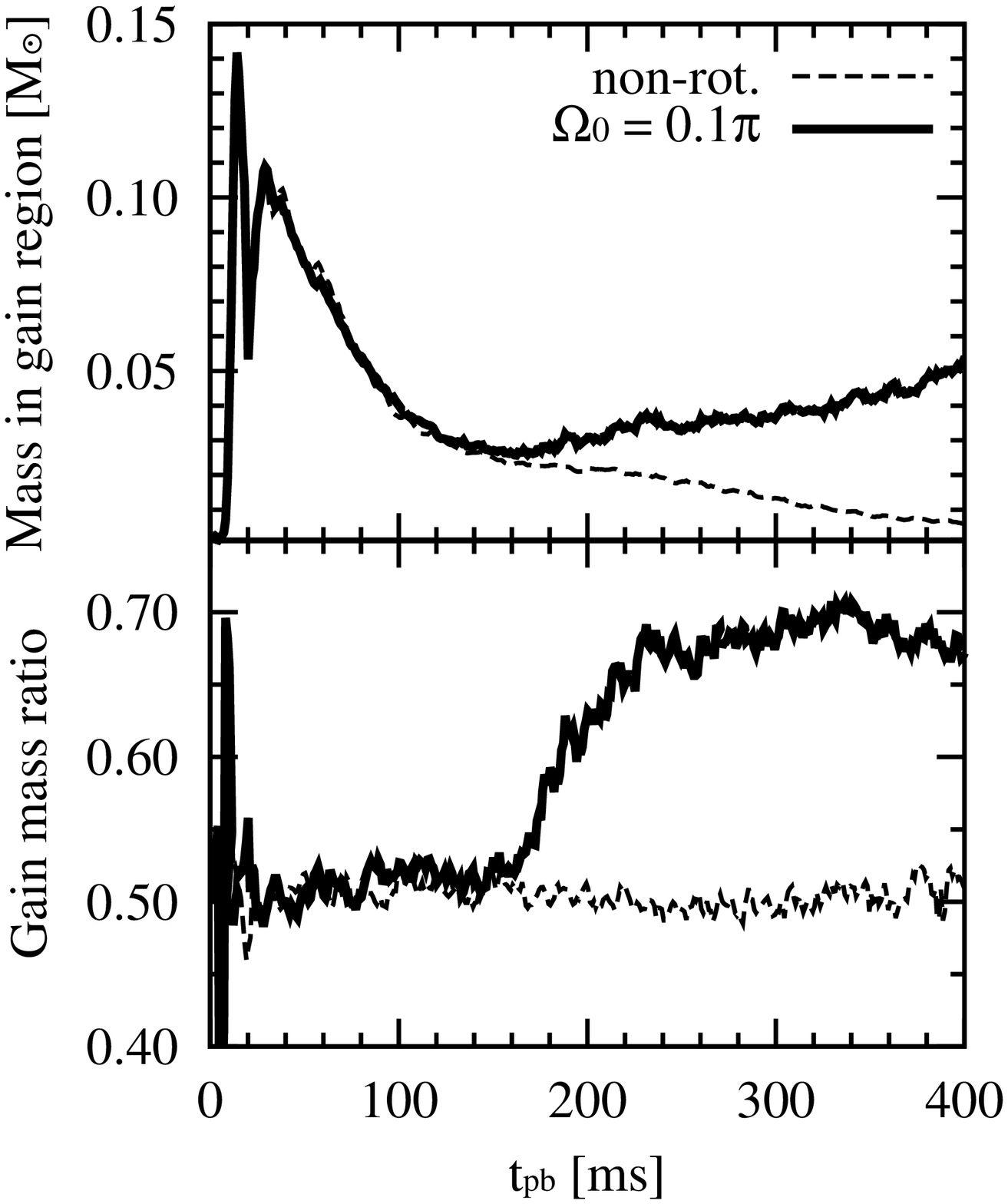}
\caption{Evolution of the mass in the gain region 
for the $L_{\nu,52} = 2.5$ model with $\Omega_0=0.0$ (dashed thin line) and 
$\Omega_0 = 0.1\,\pi$ (solid thick line). 
Top and bottom panel shows the total 
mass in the gain region and the gain-mass ratio, respectively, 
 the latter of which is defined to be the mass fraction of the 
 gain region in the range of $60^\circ \leq \theta \leq 120^\circ$ to the 
 total mass (see the text for more detail).}
\label{fig-mg25}
\end{center}
\end{figure}

From Figure \ref{fig-mg25}, one can also see that the gain mass
 of the rotating model (solid line in the top panel),
 which is almost identical to that of the non-rotating model (dashed line) 
until $t_{\rm pb} \lesssim 140 - 160$ ms, gradually increases later on.
Due to the centrifugal forces, the postshock regions deform to be 
oblate (e.g., middle left panel of Figure \ref{fig-isent}).
 This acts to enlarge the gain region in the equatorial region.
 Note in the bottom
 panel of Figure \ref{fig-mg25} that the gain-mass fraction for the 
 non-rotating model (dashed line) is close
 to 0.5 in the equatorial belt (over $60^\circ \leq \theta \leq 120^\circ$ with 
$\theta$ representing the polar angles) as it should be because the gain 
mass is randomly distributed (see again left panels in Figure \ref{fig-mlwmg25}). 
For the rotating model, the ratio exceeds 0.5, which clear shows the 
excess of the gain mass in the equatorial region. 

\begin{figure*}[tbp]
\epsscale{1.11}
\plottwo{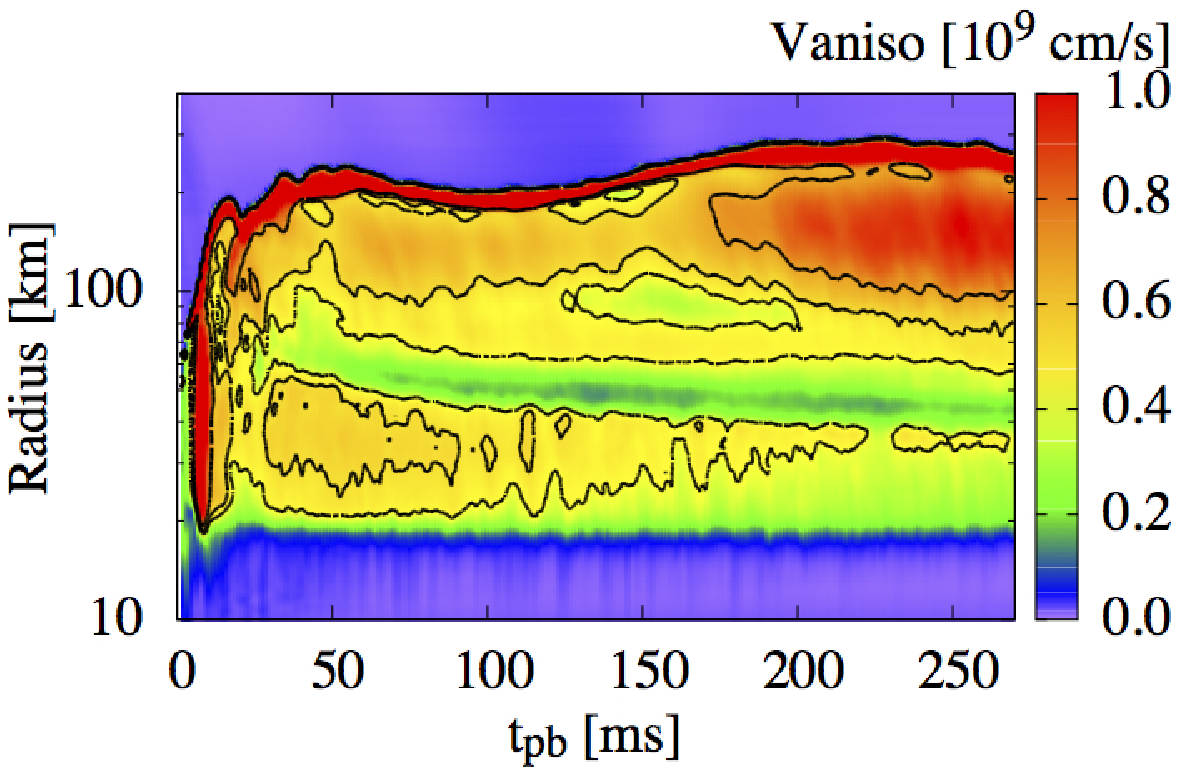}{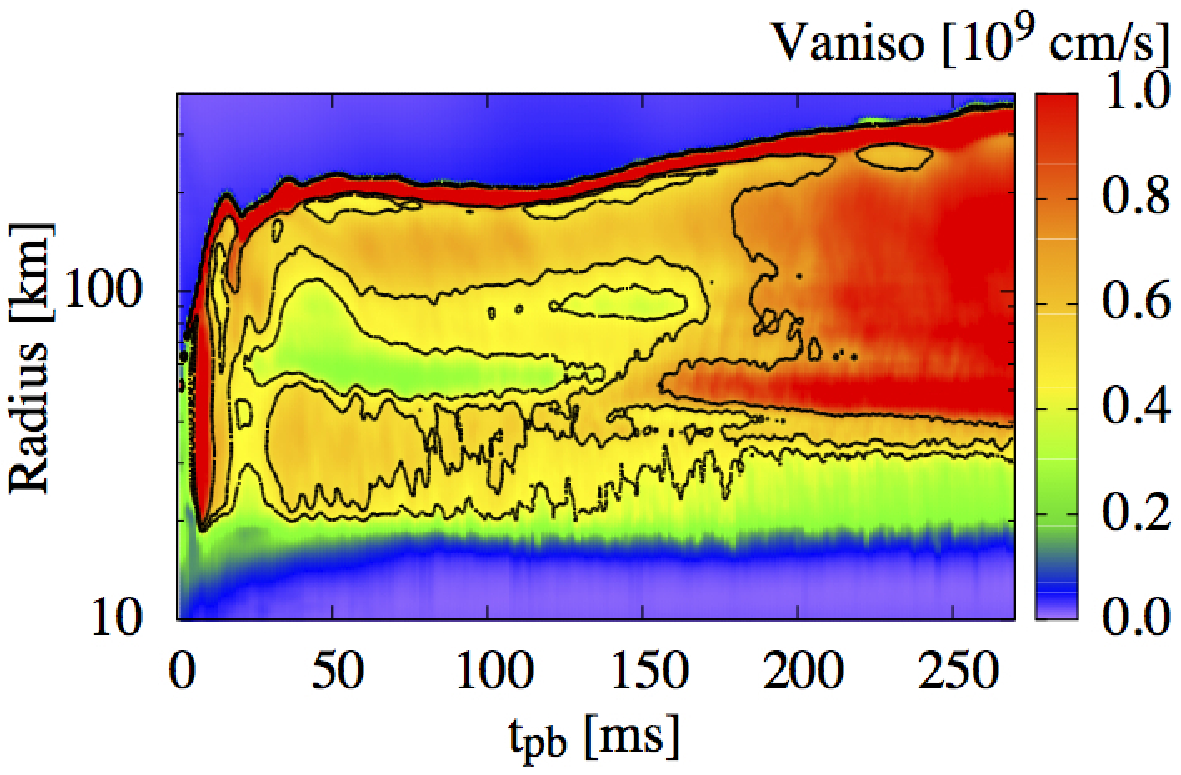}
\caption{Radial distribution of anisotropic velocity in unit of $10^9$ cm s$^{-1}$.
Two cases of $L_{\nu, 52}=2.5$ without rotation (left) and with 
$\Omega_0 = 0.1~\pi$ (right)
are shown as functions of the postbounce time, respectively.}
\label{aniso}
\end{figure*}

Figure \ref{aniso} shows the time evolution of anisotropic velocity \citep{Takiwaki12}
 which we slightly change the definition to fit to the analysis of 
rotating models as 
\begin{equation}
 v_{\rm aniso} = 
 \sqrt{ 
 \langle \rho \left(
 (v_r - \langle v_r \rangle)^2 + v_\theta^2 +  (v_\phi - \langle v_\phi \rangle)^2
 \right) \rangle
 / \langle \rho \rangle
 },
\label{eq-vaniso}
\end{equation}
in which $\langle \rangle$ denotes an average value over angles 
perpendicular to the radial direction\footnote{As a side-remark, the area of 
 intermediate anisotropic velocity (greenish regions 
at $t_{\rm pb} \lesssim 150$ ms postbounce) is slightly smaller 
for the rotating model (right panel) than for the non-rotating model 
(left panel). This may be because rotation stabilizes convection motions, 
although the effect cannot be seen 
so clearly as  
in the previous work 
\citep[e.g.,][]{fryer00,Ott08}
 simply because our models have much smaller
 initial angular momentum.}.
As seen from the right panel of 
Figure \ref{aniso}, matter with high anisotropic motions (red regions)
 starts to move outward at $t_{\rm pb} \sim 140$-$160$ ms for the rotating 
model. This is closely correlated with the time when 
the gain mass in the equatorial region starts to gradually increase (Figure \ref{fig-mg25}).
 The growth of such highly convective regions also 
 works to enhance the efficiency of neutrino heating because it makes
 longer the residence time of accreting matter in the gain region.
A brief summarize in this subsection is that the rotational flattening 
of the postshock region and the resulting increase of the total gain mass 
is primarily important to understand the reason of easier explosions of
 our 3D rotating models.

\subsubsection{Mode Analysis: SASI versus Neutrino-driven Convection}

\begin{figure}[htbp]
\begin{center}
\includegraphics[scale=0.5]{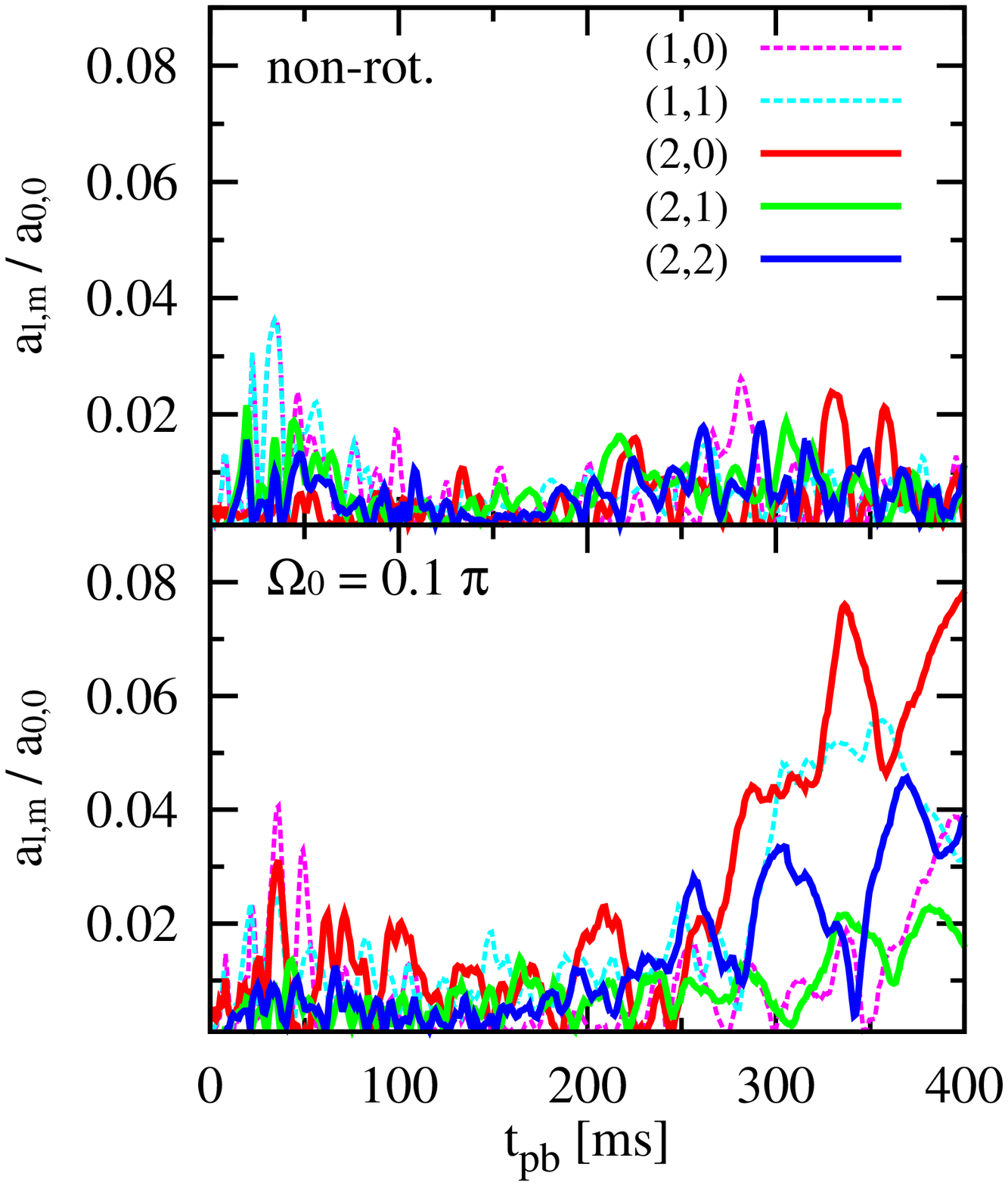}
\caption{Time evolution of the normalized spherical 
harmonic mode amplitudes ($Y_{\ell,m}$) for each set of (${\ell,m}$) 
for the $L_{\nu,52} = 2.5$ model
 without rotation (top) and with moderate 
 rotation ($\Omega_0 = 0.1\,\pi$, bottom).}
\label{fig-sasi25}
\end{center}
\end{figure}

Now we move on to perform a mode analysis of the shock 
regarding the reference models 
in the last secsion (Figure \ref{fig-sasi25}).
Comparing with the top to the bottom panel, 
it can be seen that the sloshing mode of $(\ell,m) = (2,0)$ (red line) 
for the rotating model is clearly bigger than that of the non-rotating model 
before the onset of explosion ($t_{\rm pb} \sim 140$ ms). 
As previously identified, this dipolar deformation 
(especially before the onset of explosions) is 
a typical feature of the rotating core collapse and bounce 
\citep[e.g.,][]{Moenchmeyer91,Zwerger97,Dimmelmeier02,Kotake03,Kotake04,Ott04,Ott07_prl,Dimmelmeier07,kuroda13}. 
After the stalled shock revives and turns to expansion 
($t_{\rm pb} \gtrsim 160$ ms), the dipolar axisymmetric mode 
($(\ell,m) = (2,0)$) has the largest 
amplitude, which is followed in order by low $\ell$ modes with 
non-axisymmetric deformations of $(\ell,m) = (1,1), (2,2)$, and $(2,1)$. 
Can this low-mode deformation of the shock 
be interpreted as a result of the SASI or neutrino-driven 
convection\footnote{See, for example, \citet{hanke11}, \citet{dolence12}, \citet{burrows12}, \citet{Hanke13}, and \citet{fern13} 
for a hot debate on this topic.} ?

\begin{figure*}[htbp]
\begin{center}
\plottwo{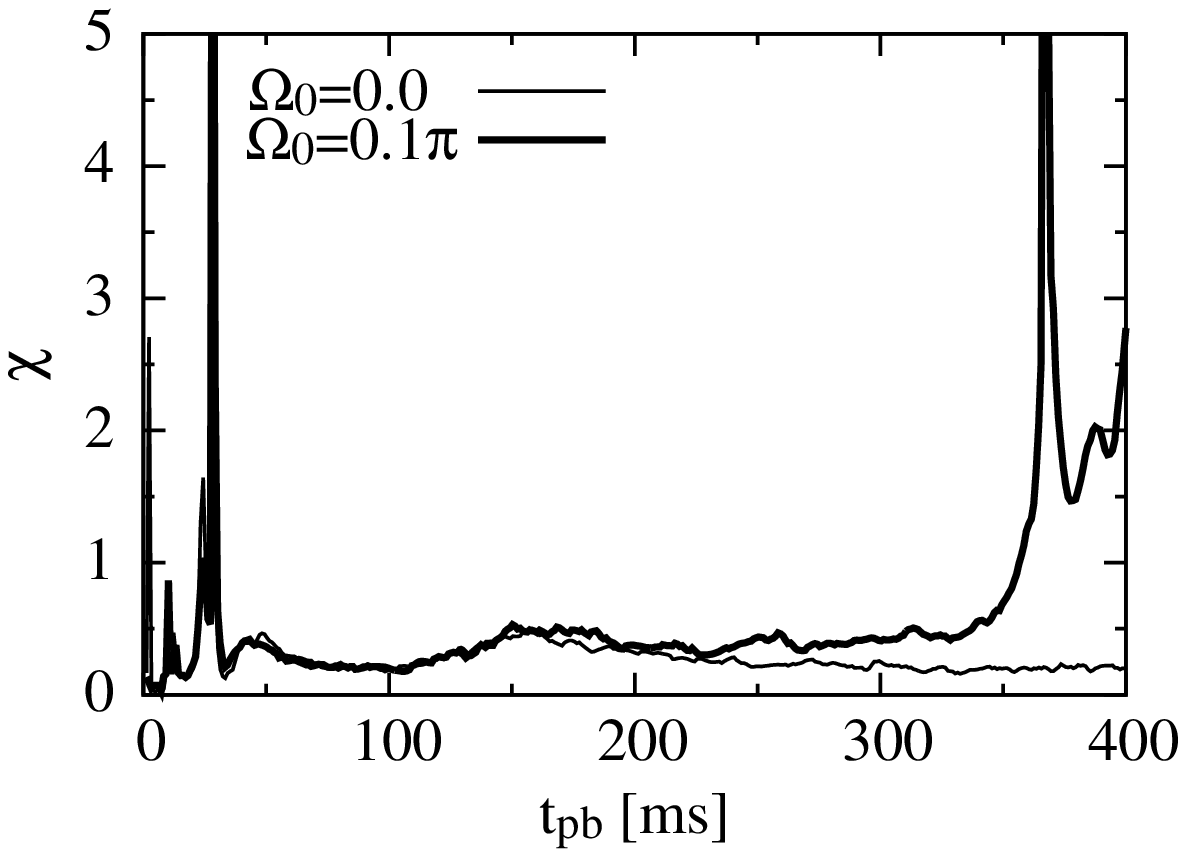}{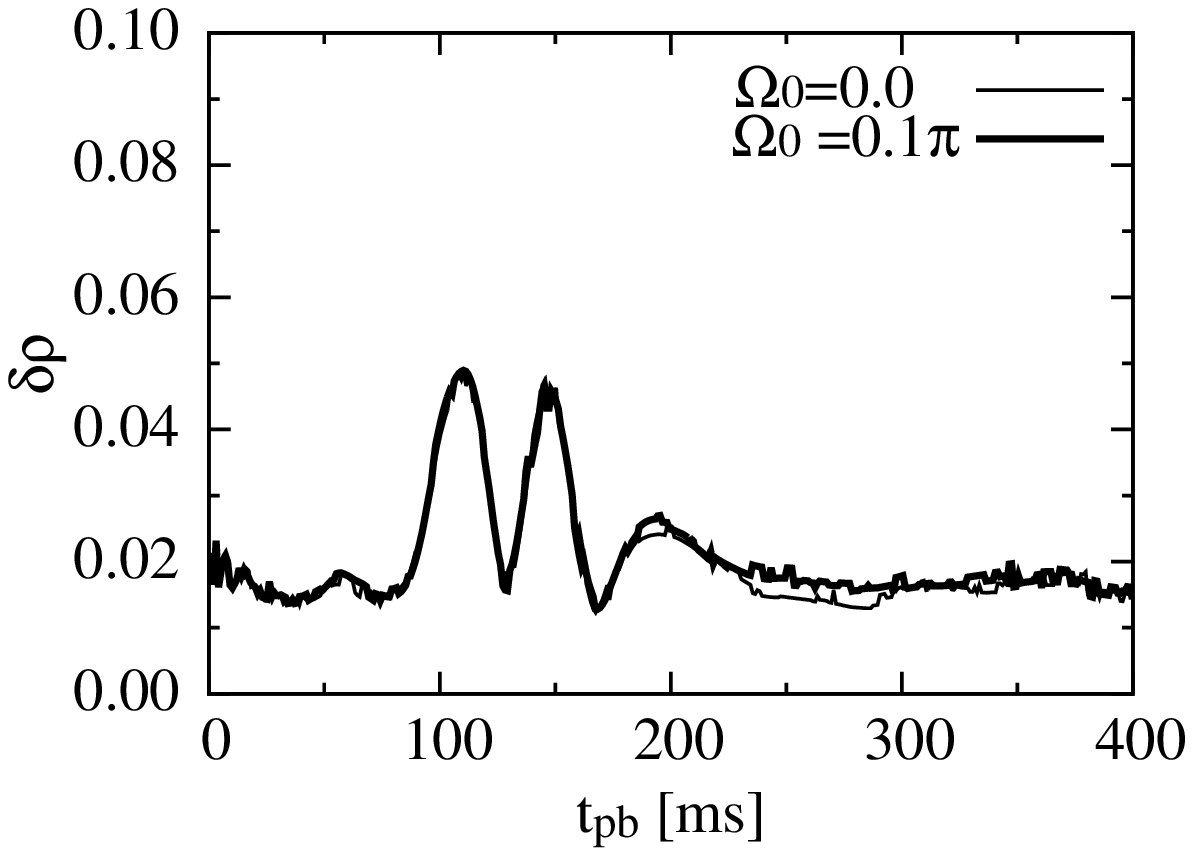}
\caption{The Foglizzo parameter (left plot) and the density perturbation
estimated at the (angle-averaged) shock surface (right plot)
as a function of postbounce time for the $L_{\nu, 52} = 2.5$ model
without rotation (thin line) and with moderate rotation 
($\Omega_0 = 0.1\,\pi$, thick line), respectively.}
\label{fig-chi25}
\end{center}
\end{figure*}

To get a hint about this question, we plot in the left panel of Figure \ref{fig-chi25} 
the Foglizzo parameter $\chi$ \citep{Foglizzo06},
the ratio of the advection to the local buoyancy 
timescale, which is evaluated as
\begin{equation}
\chi \equiv \int_{R_{\rm gain}}^{R_{\rm sh}} \omega_{\rm BV}\, \frac{dr}{|v_r|}, 
\label{eq-chi}
\end{equation}
where $R_{\rm gain}$ and $R_{\rm sh}$ denotes the angle-averaged 
gain radius and the shock radius, and 
$\omega_{\rm BV}$ is the Brunt-${\rm V\ddot{a}is\ddot{a}la}$ frequency 
\citep[e.g.,][]{Buras06a}. 
For the linear SASI regime, \citet{Foglizzo06}
found the threshold condition $\chi \gtrsim 3$ for convective
activity to develop in the gain layer. At the same time, the authors
cautioned that neutrino-driven convection becomes the primary and dominant 
instability even for $\chi \lesssim 3$ if sufficiently large 
($\gtrsim 1$\%) seed perturbations from spherical symmetry are present 
in the preshock region. These features have been already confirmed 
in both 2D and 3D simulations by
\citet{Scheck08}, \citet{burrows12}, and \citet{Hanke13}.

As shown in the left panel of Figure \ref{fig-chi25}, the $\chi \gtrsim 3$ condition
 is not generally fulfilled in any of our 3D models. However, we have 
 already seen several evidences of neutrino-driven convection such as 
in Figures \ref{fig-snap25}, \ref{fig-snap2729}, and \ref{aniso}. 
This is not surprising because as shown in the right 
panel of Figure \ref{fig-chi25} our use of the Cartesian coordinates, 
which is inferior to the use of the spherical coordinates to keep
 sphericity of the preshock region 
\citep[see extensive discussions in][]{Ott12a}, 
produces percent levels of the density perturbations there.

\subsubsection{Buoyancy-driven Shock Evolution into Explosion}\label{sec-late}
\begin{figure}[htbp]
\begin{center}
\includegraphics[scale=0.65]{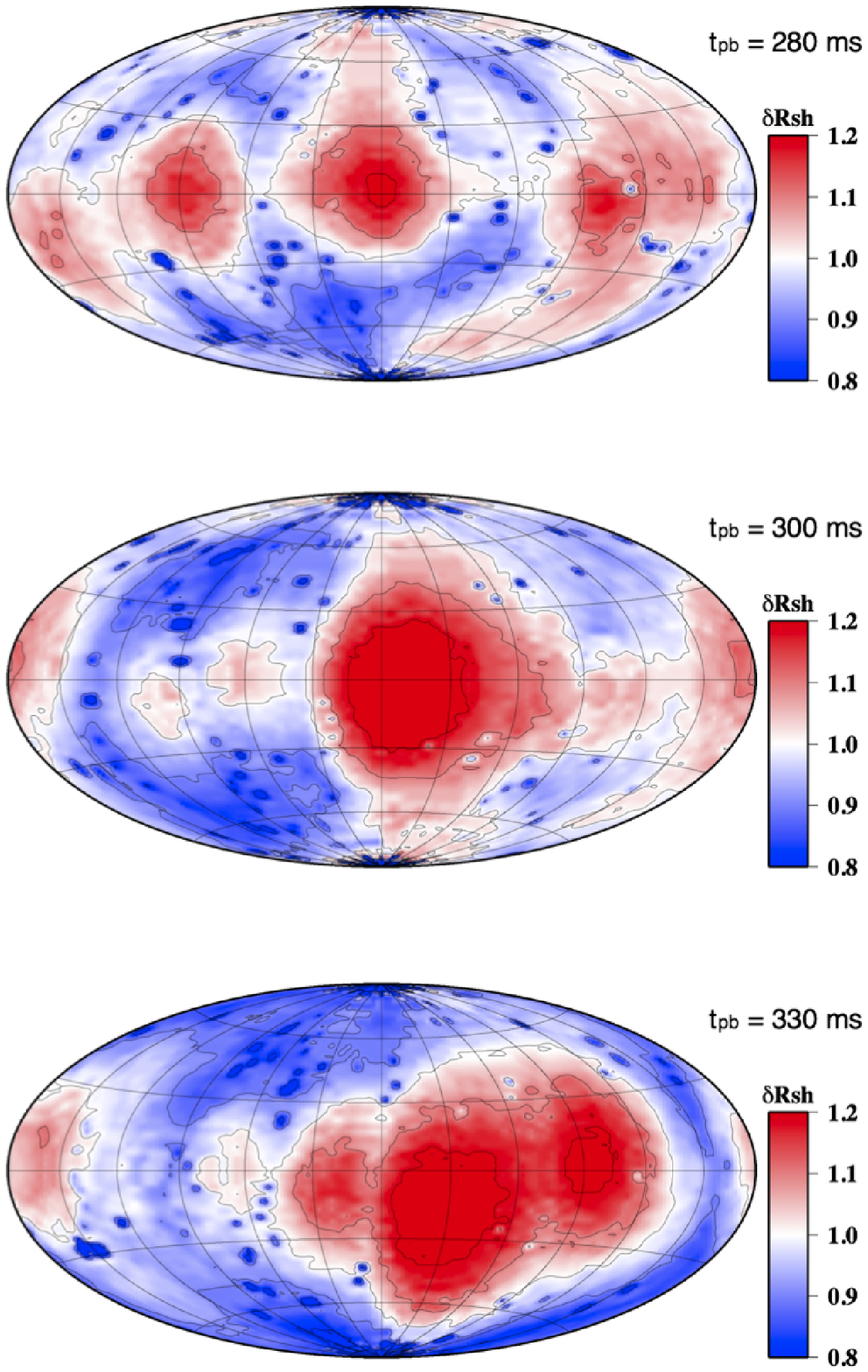}
\caption{Mollweide map of shock radius deviation from an average for the rotating model 
($L_{\nu,52} = 2.5$, $\Omega_0 = 0.1 \pi$). 
The degree of the deviation $\delta R_{\rm sh}$ is defined as 
$\delta R_{\rm sh} = R_{\rm sh}(\theta, \phi)/ \langle R_{\rm sh} \rangle$.
}
\label{fig-mlwrs}
\end{center}
\end{figure}

\begin{figure*}[htbp]
\begin{center}
\includegraphics[scale=1.]{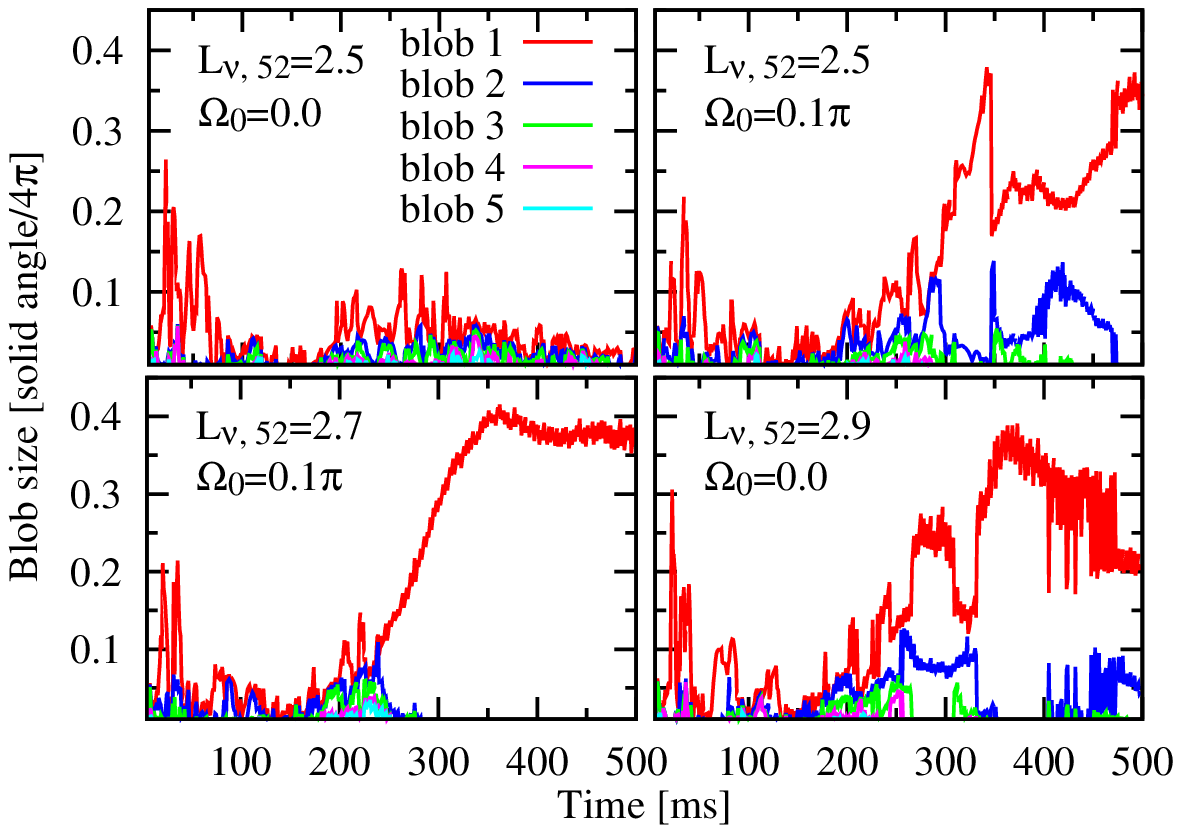}
\caption{Time evolution of blob sizes for four models 
with parameters labeled in each panel. 
Shown are solid angle fractions of the five largest blobs, 
the largest one in red and the others in different colors.
The blob sizes of a non-explosion model (top left panel) are small, 
whereas the other explosion models have large-scale blobs.
}
\label{fig-blob}
\end{center}
\end{figure*}

As mentioned above, our results 
indicate that neutrino-driven buoyant 
convection is the dominant instability that triggers the onset 
of explosion of our parameterized 3D models. All of these models show 
a monotonic shock expansion after the shock 
reaches to a certain radius ($\sim 400$ km) and does not return
 later on. In 
this section, we discuss how rotation impacts the
 buoyancy-driven convection, paying particular attention to the 
size of buoyant bubbles as in 
\citet{dolence12}, \citet{burrows12}, \citet{couch13}, \citet{murphy13} and \citet{fern13}.

As a guide to see the evolution of the buoyant bubbles,
we plot in Figure \ref{fig-mlwrs} the deviation 
 of the local shock position (defined as $\delta R_{\rm sh}(\theta,\phi)$)
from the angle-averaged value ($R_{\rm sh}(\theta,\phi)$) for our 
reference rotating model.
  From the top panel, several protrudent blobs (red regions) can be 
seen around the equator at 
 $t_{\rm pb} = 280$ ms when the shock reaches an average radius of 400 km.
Then, one of the high-entropy plumes
(located near the center of the $4\pi$ map in the middle panel) 
 rise all the way, locally pushing the shock ever larger
 radii until the global explosion is triggered
 (see the bottom panel). This feature was 
 already noted by \citet{dolence12} and \cite{couch13} in their 3D models
 for non-rotating progenitors. In this work, we furthermore point out that
 the growth of such large-scale bubbles is easier to develop 
especially in the equatorial region for 3D models that include 
(even) moderate rotation.

For a more quantitative discussion, we plot in Figure \ref{fig-blob} 
the evolution of the blob size as a function of the postbounce time.
Note in the plot that each of the blob size 
is estimated as its solid angle measured from the center. 
It is clearly shown in the plot that 
the rotating model (top right panel) 
has one large blob (red line) after the shock revival ($t_{\rm pb} \gtrsim 
140$-$160$ ms). On the other hand,
 the corresponding non-rotating model (top left panel) 
 that fails to produce an explosion
 has much smaller blobs.
In Figure \ref{fig-blob}, we add two more models with different input 
neutrino luminosity that are trending towards explosion, 
one is rotating (bottom left panel) and the 
other is non-rotating (bottom right panel).
Both of these exploding models have large plumes, which 
 continues to cover $\sim 30$-$40$ \% of the whole solid angle 
until the globally asymmetric explosions occur. 
 
This feature is commonly observed for models that exhibit the 
 runaway growth of the big bubbles.
When one of the blobs is stimulated by a stochastic factor 
such as neutrino-induced convective motion and expanding to outward, 
it is exposed to a small amount of accreting matter 
with low ram pressure.
As a result, the blob expands faster and farther than the other blobs,
 as previously pointed out by \citet{couch13} and \cite{fern13}.
Some models present very unique structures of the shock front, 
as seen in Figures \ref{fig-snap25}, \ref{fig-snap2729} and \ref{fig-isent}, 
which are far away from spherical symmetry.
It is uncertain whether these characteristic structures are maintained 
until the shock breaks out of the stellar surface.
The distorted shock front would be rounded off 
during the shock passage before the shock breakout 
or would be sharpened by nuclear energy deposition behind
 the shock \citep[see][]{nakamura12}. 
To tackle with these topics, a much longer-term, better 
 resolved 3D models covering the whole progenitor star are needed
 (see e.g., \citet{kifo03,kifo06,kifo10,handy,martin14}, and \citet{Kotake12} for 
a review), toward which 
we have attempted to make the very first step in this study.

\section{CONCLUSION AND DISCUSSION}
\label{sec:conclusion}
We performed a series of 
simplified numerical experiments to explore how rotation impacts 
the 3D hydrodynamics of neutrino-driven core-collapse supernovae.
 For the sake of our systematic study,
 we employed a light-bulb scheme to trigger explosions
 and a neutrino leakage scheme to treat 
deleptonization effects and neutrino losses from the PNS interior.
 Using a $15 \Msun$ 
progenitor, we computed thirty 3D models with a suite of the initial 
angular momentum and the light-bulb neutrino luminosity.
We find that rotation can help the onset of neutrino-driven 
explosions for models, in which the initial angular momentum is matched 
 to the one obtained in recent stellar evolutionary calculations.
 For models with larger initial angular momentum, 
 the PNS and the shock surface deform to be more oblate 
due to the larger centrifugal forces. This makes not only the gain region 
 much more concentrated around the equatorial plane, but also 
 the mass in the gain region bigger. As a result, hot bubbles 
tend to be coherently formed in the equatorial region, which pushes
 the shock ever larger radii until the global explosion is triggered.
We found that these are the main reasons that the preferred 
direction of explosion in the 3D rotating models is often perpendicular to 
the spin axis, which is in sharp contrast to the polar explosions 
around the axis that was obtained in 
previous 2D simulations.

\begin{figure*}[htbp]
\plottwo{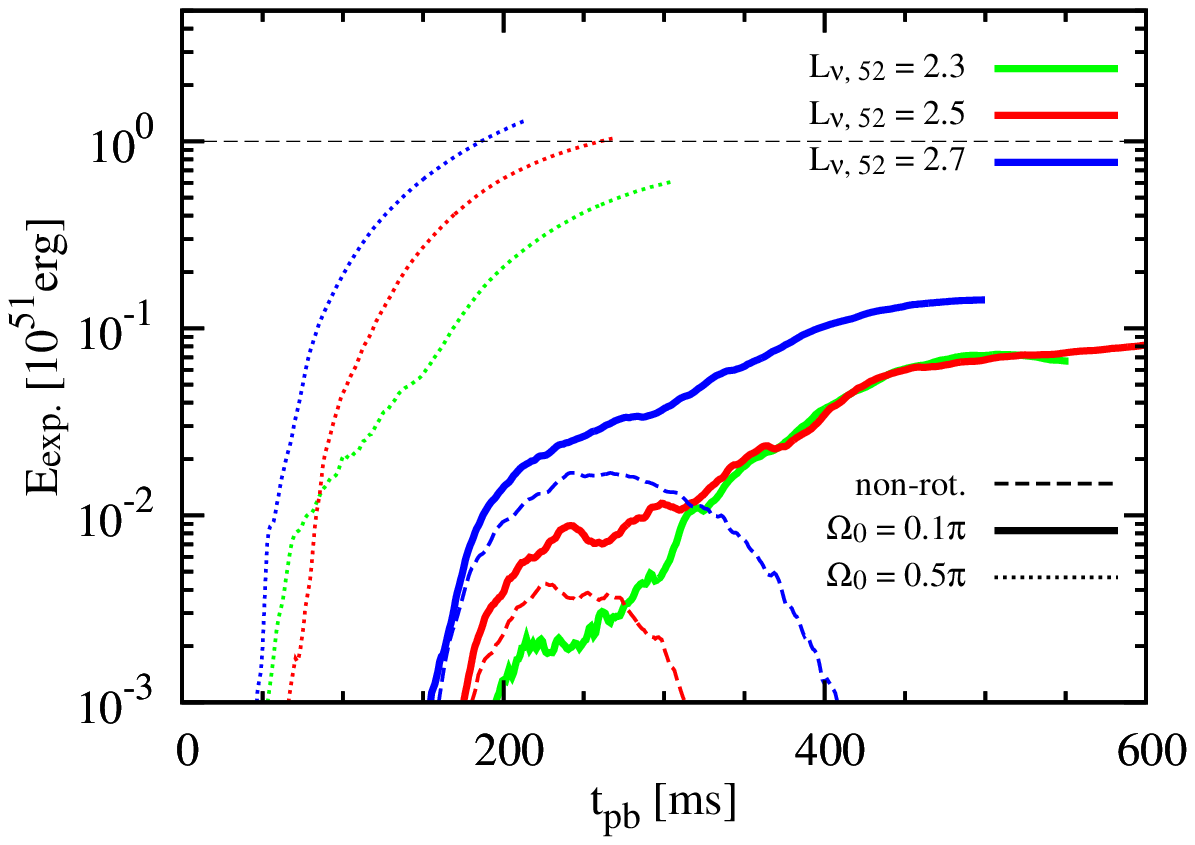}{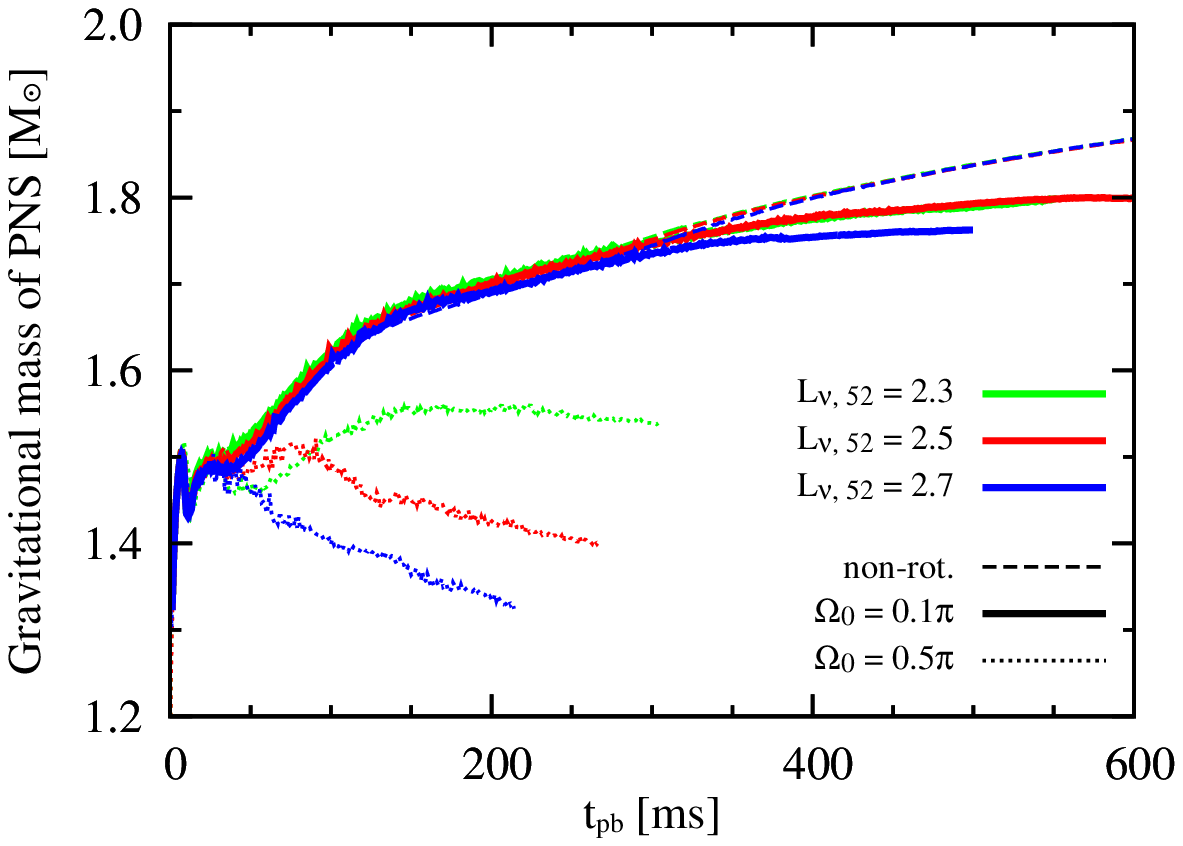}
\caption{Time evolution of diagnostic energy (left panel) 
 and the mass of the PNS (defined by a fiducial density of 
 $10^{11}~{\rm g}\,{\rm cm}^{-3}$) for the 
$L_{\nu,52} = 2.3$ (green lines), $L_{\nu,52} = 2.5$ (red), and $2.7$ (blue)
 models with different initial rotation rates.
For these three values of $L_{\nu,52}$, 
non-rotating models (dashed lines) do not explode within our simulation time, 
while moderately rotating (solid) and rapidly-rotating (dotted) models 
are trending towards explosion.
As a reference, horizontal line represents $10^{51}$ erg in the left panel. }
\label{fig16}
\end{figure*}

Finally, as a guide to discuss the strength of our parameterized explosion 
 models, we estimate a {\it diagnostic}
 energy\footnote{As in \citet{suwa10}, diagnostic energy is 
 defined as the integral of the energy all over zones that have a 
positive sum of the specific internal, gravitational, and kinetic energy.} 
and the mass of the PNS (Figure \ref{fig16}). It can be seen that our rapidly rotating models
(dotted lines with $\Omega_0 = 0.5\,\pi$, left panel)
 trending towards an energetic explosion ($\sim 10^{51}$ erg) would possibly 
leave behind a relatively reasonable remnant mass ($\sim 1.3$-$1.4 \Msun$)
for some particular choice 
of the input luminosity (e.g., dotted green line in the right panel). 
One has to be careful in interpreting this, 
because the core neutrino 
luminosity was assumed to be constant with time and spatially isotropic in 
this study. It has been shown that neutrino emission from (rapidly)
 rotating core is not isotropic at all \citep{Janka89_226,Kotake03,Ott08,brandt10},
 and more importantly, the neutrino luminosity becomes smaller for models with 
 larger initial angular momentum \citep{Ott08}. Having admitted that the predictive power of our 3D parameterized models 
 especially with rapid rotation is very limited, our 3D models 
demonstrate that even a {\it moderate} rotation ($\Omega_0 = 0.1\,\pi$) as previously thought, could significantly effect the evolution of the diagnostic 
energy and the remnant mass (compare solid with dashed line in Figure \ref{fig16}). 
A self-consistent 3D model 
\citep[ultimately with 6D Boltzmann transport, e.g.,][]{sumiyoshi12,sumiyoshi14}
is 
apparently needed to have the final word whether rotation 
will or will not lead to easier onset of neutrino-driven explosions,
 which we are going to study as a sequel of this work 
(Takiwaki et al. in preparation).

\acknowledgments
T.T. and K.K. are thankful to K. Sato for continuing encouragements.
We also thank S. Yamada, M.Liebend\"orfer, and F.K. Thielemann 
for stimulating discussions.
Numerical computations were carried out in part on 
XC30 and general common use computer system at the center for 
Computational Astrophysics, CfCA, 
the National Astronomical Observatory of Japan,
Oakleaf FX10 at Supercomputing Division in University of Tokyo, and on 
SR16000 at YITP in Kyoto University. This study was supported in part 
by the Grants-in-Aid for the Scientific Research 
from the Ministry of Education, Science and Culture of Japan (Nos. 20740150,
23540323, 23340069, 24244036, and 26870823) and by HPCI Strategic Program of 
Japanese MEXT.


\end{document}